\documentclass[fleqn,usenatbib]{mnras}

\usepackage{newtxtext,newtxmath}

\usepackage[T1]{fontenc}

\DeclareRobustCommand{\VAN}[3]{#2}
\let\VANthebibliography\thebibliography
\def\thebibliography{\DeclareRobustCommand{\VAN}[3]{##3}\VANthebibliography}

\usepackage{graphicx}	
\usepackage{amsmath}	
\usepackage{multirow}   
\usepackage{booktabs}   
\usepackage[flushleft]{threeparttable} 

\title[The Fate of Globular Cluster Substructure]{The Fate of Globular Cluster Substructure: A Kinematic Response to Galaxy Assembly}

\author[F. A. Pal et al.]{
Finn A. Pal,$^{1}$\thanks{E-mail: f.pal@unsw.edu.au}
Sarah L. Martell,$^{1}$
Elizabeth J. Iles$^{1,2}$
\\
$^{1}$School of Physics, University of New South Wales, NSW, 2052, Australia\\
$^{2}$Sydney Institute for Astronomy (SIfA), Sydney University, NSW, 2050, Australia
}

\date{Accepted XXX. Received YYY; in original form ZZZ}

\pubyear{\the\year{}}

\begin{document}
\label{firstpage}
\pagerange{\pageref{firstpage}--\pageref{lastpage}}
\maketitle

\begin{abstract}
Globular clusters (GCs) are powerful tracers of galaxy assembly, frequently used to identify accreted substructure and reconstruct hierarchical merger histories. With advances in GC formation models and cosmological simulations, we can now better quantify the information about galaxy evolution encoded in present-day GCs. Here, we investigate how GC kinematics evolve over cosmic time and assess the extent to which GCs retain memory of the past of their host galaxy. Using a GC formation model applied to five Milky Way (MW) analogues from the Latte suite of the FIRE-2 simulations, we track the evolution of kinematic properties. At $z=0$, \textit{in-situ} and \textit{ex-situ} GCs exhibit substantial overlap in kinematic space, indicating that these populations are not clearly separable. We find that a subset of kinematic properties evolve in an ordered fashion across both \textit{in-situ} and \textit{ex-situ} populations, whereas others are dominated by stochastic variations. As a result, by the present day, most memory of the progenitor of an accreted GC is erased and only a few correlations persist. These correlations link progenitor halo mass to the total mass and number of a GC population, and the galactocentric distance of GC substructure to progenitor maximum circular velocity. These results highlight how both deterministic and stochastic processes driven by galaxy evolution shape GC kinematics and demonstrate the limits of reconstructing the assembly history of a galaxy from present-day GC orbits alone.
\end{abstract}




\section{Introduction}\label{sec:introduction}
In our current understanding of galaxy formation and evolution the Lambda Cold Dark Matter ($\Lambda$CDM) model dictates that the large-scale structures we observe today are the result of hierarchical mergers \citep{White_1978, Blumenthal_1984}. As such, being able to identify any structure within a galaxy that may be a remnant of a past merger can help unlock the pathways through which it has evolved.
The Milky Way (MW) provides an ideal location to search for this type of structure, with streams such as Sagittarius \citep{Ibata_1994}, Helmi \citep{Helmi_1999}, Cetus \citep{Yuan_2019} and Wukong \citep{Yuan_2020} providing detail on the physics driving the process of accretion. The advent of the Gaia mission \citep{Gaia_2016, Gaia_2018, Gaia_2023} has also introduced a wealth of new kinematic information that has enabled the reconstruction of stellar orbits within the MW. In combination with abundance information from spectroscopic surveys such as GALAH \citep{De_Silva_2015}, APOGEE \citep{Majewski_2017}, LAMOST \citep{Zhao_2012}, the H3 survey \citep{Conroy_2019} and S$^5$ \citep{T_S_Li_2019}, to name a few, there has been the identification of substructure in chemo-dynamical phase space that is believed to be linked to ancient mergers. This includes the identification of the Gaia Sausage/Enceladus \citep[GS/E; ][]{Helmi_2018, Belokurov_2018, Haywood_2018} which is believed to have been the MW's last major merger, accreted approximately 8-11 Gyr ago. Additional chemo-dynamical substructures that have also been interpreted as evidence of past accretion include Thamnos \citep{Koppelman_2019}, Sequoia \citep{Myeong_2019} and Heracles \citep{Horta_2021}. However, not all identified substructures are necessarily accreted. For example, the Aurora population \citep{Belokurov_2022} has instead been interpreted as a reflection of the chaotic pre-disc era of the MW. Recognising these features has enabled Galactic archaeologists to not only piece together different stages of the MW's evolution but has also allowed for the comparison of current streams to ancient merger substructure and the development of a timeline for the process of accretion.

This search for evidence of past accretion events goes beyond just the grouping of field stars to include globular clusters (GCs). Although the origin and evolution of GC systems is still a topic of investigation \citep[e.g.][]{Forbes_2018_feb}, there has been an ongoing effort to use GC properties to distinguish clusters which formed within the main progenitor of the MW (\textit{in-situ}) from those which have been accreted (\textit{ex-situ}) \citep[e.g.][]{Belokurov_2023, Belokurov_2024}. There has been a particular emphasis placed on clustering GCs by their orbital integrals of motion (IoMs) and chemical abundances \citep[e.g.][]{Massari_2019, Callingham_2022, Malhan_2022, Chen_2024_mar, Massari_2025, De_Leo_2025} and searching for distinct tracks in age-metallicity relations \citep[][]{Leaman_2013, Kruijssen_2019, Forbes_2020} to help distinguish individual accretion events. 

The major challenges in identifying these different accretion events are: that we can only rely on observations of GCs' current properties and orbits, and that the allocation of MW GCs to different accretion events can be impacted by the method and/or dataset employed. While there is consensus in the classification of some GCs, there still remain discrepancies in others \citep[for an example comparison of classifications, see][]{Chen_2024_mar}. The complexity of this task is underscored by the estimate that the MW has accreted roughly 100 satellite galaxies over its history \citep{Bland_Hawthorn_2016}. Although not every accretion event is expected to contribute a GC that survives to $z=0$, this nonetheless highlights the inherent difficulty of assigning progenitors. For some GCs, allocation to an accretion event is straightforward, particularly when they can be spatially associated with known Galactic streams, such as those from Sagittarius \citep{Bellazzini_2020}. In other cases, studies like \citet{Callingham_2022} have highlighted that mixing across parameter spaces can blur distinctions, making it challenging to separate GCs from different progenitors. In the current scheme of identifying accreted substructure, there is an assumption that \textit{ex-situ} GCs retain some signal of the progenitor from which they formed, whether it be kinematic, spatial or chemical. However, simulation work presented in \citet{Jean_Baptiste_2017, Koppelman_2020} and \citet{Arora_2022} have demonstrated that mergers can cause significant kinematic mixing, such that the remnants of an accreted satellite may no longer form a single, clearly defined over-density in kinematic space. Furthermore, the development of structure, such as bars \citep[e.g.][]{Price_Whelan_2016, Dillamore_2024} and spiral arms \citep[e.g.][]{Sellwood_2002, Arunima_2025, Della_Croce_2025}, have also been found to alter the orbital kinematics of both stars and star clusters.

Consequently, there has been a strong push to use GC simulations to test whether the substructure we identify is a genuine signature of evolution. Because of the high computational cost of directly modelling GC formation and evolution within realistic galaxies, the primary approach in cosmological simulations has been to model GCs through post-processing. In this scheme, the formation and evolution of GCs is described through analytical models and then applied to pre-existing simulations, with examples including \citet{Renaud_2017, Valenzuela_2021, De_Lucia_2024} and \citet{Chen_2024_jan}. Additional frameworks have also been developed in which cluster formation is either implemented on-the-fly or inferred from the simulated galaxy properties. One example is MOSAICS \citep{Kruijssen_2011, Kruijssen_2012}, implemented within the EAGLE project \citep{Crain_2015, Schaye_2015} and extended in E-MOSAICS \citep{Pfeffer_2018}. Another example is the EMP-Pathfinder framework \citep{Reina_Campos_2022}. In these approaches, cluster formation and evolution models are coupled to galaxy formation simulations, in which GCs are evolved on-the-fly as sub-grid components of the stellar population. This allows the properties of the cluster population to be linked directly to the evolving galaxy environment. Collectively, the variation in these models highlights the range of approaches to GC formation and evolution, with a comparative analysis of their age predictions presented in \citet{Valenzuela_2025}, providing useful context for interpreting each framework.

In this work, we adopt the GC formation model of \citet{Chen_2024_jan}, representing the latest stage of a long-term project that builds upon \citet{Muratov_2010, Li_2014, Choksi_2018, Chen_2022, Chen_2023}. Through the application of this model we examine how galaxy evolution influences the kinematic substructure of GCs. We assess the ability of clustering algorithms to distinguish GCs originating from different progenitors using purely kinematic information and investigate the evolutionary processes that drive kinematic mixing. In particular, we aim to differentiate between mechanisms that induce ordered versus stochastic changes in GC kinematics. Finally, we explore whether, despite kinematic mixing, GCs retain any kinematic signatures of their original accretion events.

This paper is organised as follows. Section~\ref{sec:method} describes the methodology adopted in this work, including the GC formation model and its coupling to cosmological simulations, the procedures used to model the gravitational potentials of simulated galaxies, the identification of disc formation, and the calculation of orbital actions. Section~\ref{sec:orbital_evolution} examines the effectiveness of clustering kinematic properties to identify GC substructure and explores how the kinematic evolution of GCs is linked to the structural evolution of their host galaxies. In Section~\ref{sec:information_loss}, we identify evolutionary phases in which deterministic or stochastic processes dominate GC kinematics, and investigate which kinematic signatures persist post-accretion and can be used to better understand past merger events. The main conclusions of this work are summarised in Section~\ref{sec:conclusion}.

\section{Method}\label{sec:method}
This study examines GC formation and evolution across five galaxies drawn from the Latte suite of the FIRE-2 simulations \citep{Wetzel_2016}. In this Section, we first summarise the FIRE-2 simulations and outline our galaxy selection. We then describe the GC formation model and its application to these systems. Finally, we discuss the dynamical modelling of the galaxies, including disc modelling, potential fitting, and the calculation of orbital actions.

\subsection{Model}\label{subsec:model}
\subsubsection{Simulation}\label{subsubsec:simulation}
We use the publicly available FIRE-2 cosmological zoom-in simulations \citep{Wetzel_2023, Wetzel_2025}, from the Feedback In Realistic Environments (FIRE) project, generated using the {\sc Gizmo} code \citep{Hopkins_2015} and the FIRE-2 physics model \citep{Hopkins_2018_oct}. We focus specifically on the Latte simulation suite from the first public data release, which presents a collection of zoom-in simulations of isolated galaxies with halo masses comparable to that of the MW \citep{Wetzel_2016, Wetzel_2023}. The first public release includes complete merger trees and halo catalogues across 601 snapshots from $z = 99$ to $z = 0$, with full particle data available for 39 snapshots starting from $z = 10$.

\begin{table}
\caption{General properties of the five FIRE-2 Latte suite galaxies at $z=0$. Mass and radius values have been taken from \citet{Wetzel_2023, Quinn_2025}.}
\label{tab:galaxy_properties}
\centering
\begin{tabular}{ccccccc}
\hline
\multirow{2}{*}{Name} &
R$_{200\text{m}}$ & M$_{200\text{m}}$ & R$_{\star, 90}$ & M$_{\star, 90}$ & \multirow{2}{*}{N$_{\text{GC}}$} & \multirow{2}{*}{N$_{\text{acc}}$} \\
[2pt]
& [kpc] & [$10^{12}$M$_{\sun}$] & [kpc] & [$10^{10}$M$_{\sun}$] & & \\ \hline
m12b & 358 & 1.43 & 10.9 & 8.5  & 234 & 4  \\
m12c & 351 & 1.35 & 10.4 & 5.8  & 237 & 2  \\
m12f & 380 & 1.71 & 13.3 & 7.9  & 250 & 6 \\
m12i & 336 & 1.18 & 10.0 & 6.3  & 211 & 5  \\
m12m & 371 & 1.58 & 12.5 & 11.0 & 324 & 5  \\ \hline
\end{tabular}
\begin{threeparttable}
\begin{tablenotes}
\item \textit{Notes.}
\item Name: Name of the simulation.
\item R$_{200\text{m}}$: Radius enclosing a mean density 200$\times$ the Universe matter density.
\item M$_{200\text{m}}$: Total mass enclosed at R$_{200\text{m}}$.
\item R$_{\star, 90}$: Radius enclosing 90\% of the stellar mass within the central 20 kpc. 
\item M$_{\star, 90}$: Total stellar mass enclosed at R$_{\star, 90}$.
\item N$_{\text{GC}}$: Average number of GCs at $z=0$ over 100 GC model iterations.
\item N$_{\text{acc}}$: Accretion events contributing $\geq 5$ GCs surviving at $z=0$.
\item Simulation references are as follows: m12b, m12c \citep{Garrison_Kimmel_2019}; m12f \citep{Garrison_Kimmel_2017}; m12i \citep{Wetzel_2016}; m12m \citep{Hopkins_2018_oct}.   
\end{tablenotes}
\end{threeparttable}
\end{table}

In the selection of galaxies for this work, we include three simulations (m12f, m12i, and m12m) which \citet{Sanderson_2020} previously demonstrated to broadly resemble the MW in terms of stellar mass, scale radius, and gas fractions at $z = 0$. We further increase our sample size and search for generalised trends across differing evolutionary paths by also including simulations m12b and m12c. Like the first three, these galaxies are MW analogues with similar mass and disc morphology, and were run using the same 'AGORA' cosmology as m12f, m12i and m12m \citep[$\Omega_{\mathrm{m}} = 0.272$, $\Omega_\Lambda = 0.728$, $\Omega_{\mathrm{b}} = 0.0455$, $h = 0.72$, $\sigma_8 = 0.807$, $n_s = 0.961$,][]{Kim_2014}, ensuring consistency in the underlying simulation parameters. Table~\ref{tab:galaxy_properties} lists the general properties of each galaxy at $z=0$, along with a citation for the corresponding simulation for further details.

\begin{figure*} 
    \centering
    \includegraphics[width=0.95\textwidth]{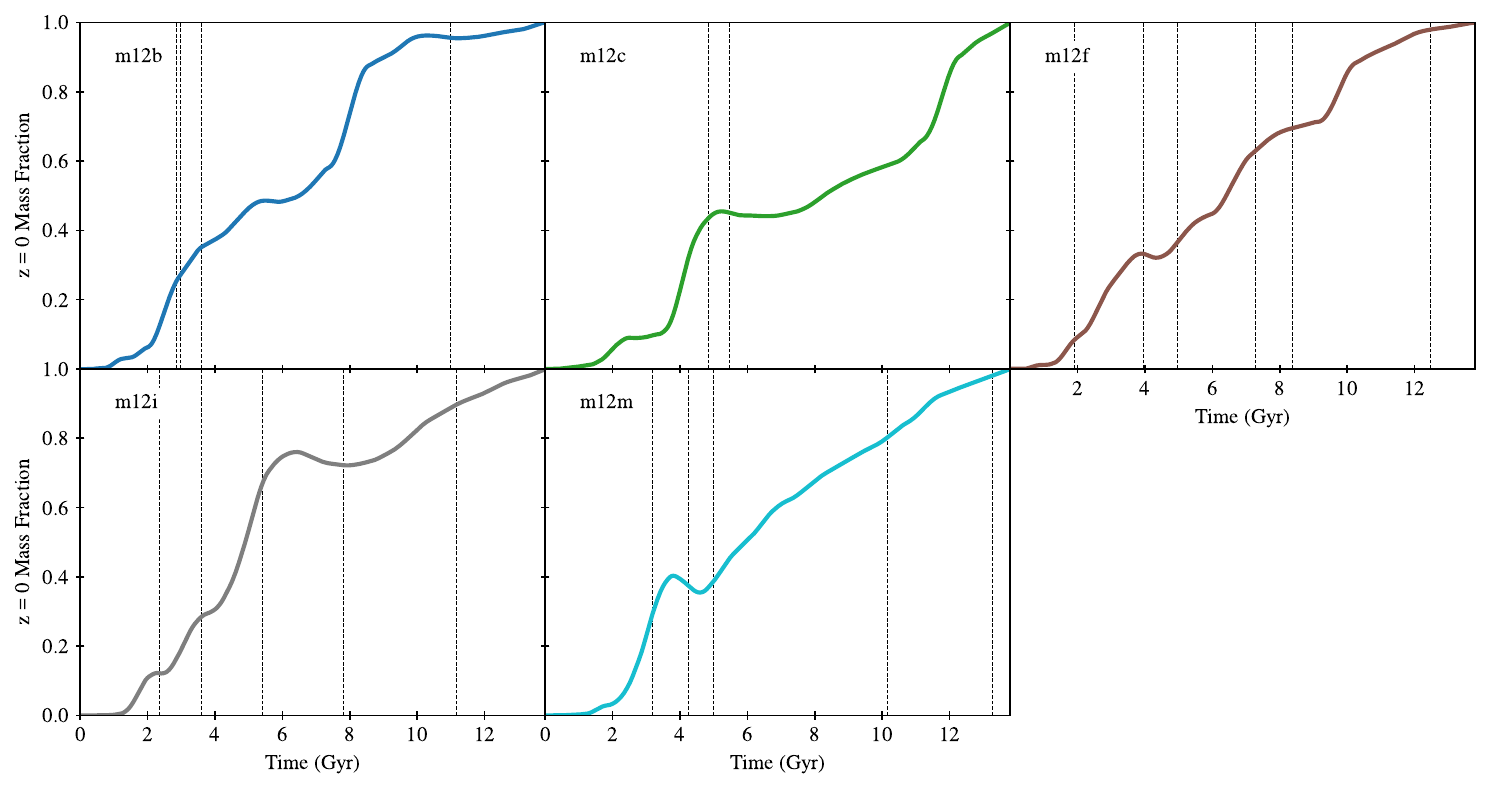}
    \caption{Mass growth history of each simulated galaxy, normalised by its mass at $z=0$. Mass growth lines have been smoothed using a local polynomial Savitzky-Golay filter. Each vertical dashed line marks an accretion event that when averaged over 100 iterations of the GC formation model, contributes at least 5 GCs to the main galaxy that survive until $z=0$. Galaxy m12f exhibits the largest number of GC accretion events (6), while m12c has the fewest (2).}
    \label{fig:mass_growth}
\end{figure*}

Across these five galaxies, baryonic (gas and star) particles each have initial masses of 7070 M$_\odot$, while dark matter (DM) particles have masses of 35,000 M$_\odot$. The simulations can resolve structures down to parsec scales, with DM and star particles assigned fixed softening lengths of $h_{\mathrm{DM}} = 20$ pc and $h_{\mathrm{star}} = 4$ pc, respectively. Gas particles use adaptive kernel/softening lengths, with a median value $h_{\mathrm{gas}} = 25-40$ pc at $z=0$ \citet{Wetzel_2016}. For FIRE galaxies, the halo catalogues include only the DM mass and so we obtain an approximation of the total (DM + baryonic) halo mass through scaling by $(1-f_{\mathrm{b}})^{-1}$, where $f_{\mathrm{b}} \equiv \Omega_{\mathrm{b}} / \Omega_{\mathrm{m}}$. We present the mass growth histories with the above correction, normalised by the $z=0$ mass in Figure~\ref{fig:mass_growth}. 

\subsubsection{GC formation model}\label{subsubsec:gc_model}
\defcitealias{Chen_2024_mar}{2024a}
In this work, we use the \citet{Chen_2024_jan} version of the GC formation model. As the model has been thoroughly detailed in \citet{Chen_2022, Chen_2023, Chen_2024_mar, Chen_2024_jan}, we provide here only a brief summary of the relevant steps.

\begin{enumerate}
    \item \textbf{Cluster formation:} When the specific accretion rate of a galaxy's DM halo satisfies the condition $\dot{M_h} / M_h > p_3$ (see Appendix~\ref{asec:specific_accretion_rate}), a GC formation event is triggered. The threshold parameter $p_3$ controls how frequently such events occur. To determine the total mass of GCs that form during such an event, a series of scaling relations \citep{Behroozi_2013, Lilly_2013, Genzel_2015, Tacconi_2018, Wang_2022} are used to infer the galaxy’s total gas mass at that time, with a combination of Gaussian process and Gaussian noise used to model the scatter in these relations. Following \citet{Kravtsov_2005}, the total mass of GCs forming is then given by $M_{\text{tot}} = 1.8 \times 10^4 p_2 M_{\text{gas}}$, where $p_2$ controls the efficiency of cluster formation from the available gas. These formation events can occur in any halo that satisfies these conditions. Clusters are considered to be \textit{in-situ} if they formed at any time in the main progenitor branch of the MW analogue, i.e., the galaxy that evolves into the $z=0$ MW analogue, and \textit{ex-situ} if they formed in other galaxies and were later accreted onto this branch.
    
    \item \textbf{Cluster sampling:} To determine the initial mass of an individual GC, we sample a Schechter function \citep{Schechter_1976} with an exponential cutoff of $10^7 M_\odot$. Clusters masses are drawn probabilistically from this function, with any mass $< 10^4 M_\odot$ discarded, as \citet{Chen_2023} indicated that clusters below this mass would be fully disrupted within 1 Gyr. Sampling continues until the total mass of newly formed clusters just exceeds $M_{\text{tot}}$.

    \item \textbf{Particle Assignment:} After the number of newly formed GCs and their initial masses have been determined, the next phase of the model is to allocate GCs to tracer particles. These particles are used track the positions and velocities of the GCs and enable the calculation of other physical properties, such as orbital actions (Section~\ref{subsubsec:orbital_actions}). Collisionless (star and DM) particles are selected as tracers, with a defined order of preference. First, newly formed star particles (age < 10 Myr) within 3 kpc of the galaxy centre \citep[approximately twice the effective radius of newly formed stars during the period of GC formation,][]{van_der_Wel_2014, Shibuya_2015, Pillepich_2019} are selected. This radial limit follows the GC-formation prescription of \citet{Chen_2022}, who restricts cluster formation to the inner regions of galaxies based on observations revealing that massive star clusters preferentially form in galactic centres \citep[e.g.][]{Adamo_2015, Adamo_2020, Randriamanakoto_2019}. We adopt the same physically motivated radial criterion to remain consistent with the underlying model. If an insufficient number of young star particles is available near the galactic centre, the model then considers older star particles, and finally resorts to DM particles if needed.

    \item \textbf{Cluster evolution:} Immediately after formation, GCs are modelled to undergo a rapid mass loss of 45\% due to stellar evolution \citep{Gieles_2023}. This is treated as an instantaneous process, justified by its much shorter timescale compared to the typical GC lifetime \citep{Chen_2024_jan}. Subsequent mass loss from tidal disruption is then modelled using a power-law function that depends on the cluster’s initial mass, current mass, and the strength of the surrounding tidal field. The tidal field is parametrised by the angular frequency ($\Omega_{\mathrm{tid}}$) and estimated from the eigenvalues of the tidal tensor. These eigenvalues are numerically approximated by evaluating the gravitational potential across a $3 \times 3 \times 3$ cubic grid of side length = 300 pc, centred on each GC particle. To improve the accuracy of this approximation, a third tunable parameter $\kappa$, is introduced to modulate the intensity of the tidal field such that $\Omega_{\mathrm{tid}}^2 = \kappa (\hat{\lambda}_1 - \hat{\lambda}_3)$. For further details, see \citet{Chen_2023, Chen_2024_jan}.
\end{enumerate}

As presented in \citet{Chen_2024_mar}, the optimal parameters for applying the GC model to the Latte suite are ($p_2$, $p_3$, $\kappa$) = (7, 1 Gyr$^{-1}$, 1.5). One complication with using the FIRE-2 simulations is that, as noted in Section~\ref{subsubsec:simulation}, the first public data release \citep{Wetzel_2023} does not include particle information for every snapshot. Consequently, cluster formation (which relies solely on merger tree information) and particle assignment (which requires both merger tree and particle data) must be performed asynchronously and then combined at snapshots where complete information is available. For further details on the asynchronous application of the GC model to FIRE snapshots, see \citet{Chen_2024_mar}.

For each of the Latte suite simulations, we run the GC formation model across 100 iterations, with each iteration representing a unique realisation of the model. Variation between iterations arises partly because the Gaussian process and Gaussian noise that represent scatter in the relevant scaling relations are re‑sampled for each iteration, producing a different total cluster mass whenever a formation event is triggered \citep{Chen_2024_mar}. Further variation is introduced in the number and individual masses of clusters, as masses are drawn probabilistically from the Schechter‑type initial cluster mass function. In the selection of tracer particles, although some of the same young star particles may be chosen across iterations, differences naturally occur when many eligible particles are available near the galactic centre. Additionally, when older star particles or DM particles are required to act as tracers, these are not necessarily identical between iterations. As a result, each iteration follows the same formation and evolutionary criteria but yields a distinct stochastic realisation of the GC population. This approach allows us to explore the range of possible outcomes of the GC model and identify robust links between GC evolution and galaxy properties.

We find that 100 iterations is enough to highlight general trends amongst the sample of galaxies. We also note that within this work, when examining \textit{ex-situ} GCs, we focus on mergers that provide at least 5 GCs at $z=0$, on average across the 100 iterations of the model. We make the assumption that any accreted substructure with GC counts below this threshold are undetectable via clustering. Merger events with more than 5 GCs remaining at $z=0$ are presented in Figure~\ref{fig:mass_growth}, with m12c having the fewest GC accretion events (2) and m12f having the most (6). 

\subsection{Dynamical Modelling}\label{subsec:dynamical_modelling}

\subsubsection{Modelling potentials}\label{subsubsec:methodling_potentials}
To model the gravitational potentials of the simulated galaxies, we make use of the galactic dynamics software {\sc agama} \citep{Vasiliev_2019}. We simulate a time-evolving potential by fitting an independent potential to each simulation snapshot based on the approach outlined in \citet{Arora_2022, Arora_2024} and summarised in this section. 

For the construction of the potential, we assume axisymmetric conditions and select all particles within 2$R_{\text{vir}}$ of the galactic centre. The DM halo and hot gas ($T_{\text{gas}} \geq 10^{4.5}$ K) components are modelled using a multipole expansion in spherical coordinates with the angular structure being represented by a sum of spherical harmonic functions in $\theta$ and $\phi$, truncated at $l_\text{max} = 4$. The radial dependence is described by arbitrary functions of radius ($r$), sampled at 20 logarithmically spaced radial nodes and interpolated using quintic splines \citep{Vasiliev_2013}. The resulting potential defining the halo is presented as:
\begin{equation}\label{eqn:phi_halo}
    \Phi_{\text{halo}}(r, \theta, \phi) = \sum^{l_\text{max}}_{l=0} \sum^{l}_{m=-l} \Phi_{lm}(r) Y^{m}_{l}(\theta, \phi).
\end{equation}
Prior to disc formation (see Section~\ref{subsubsec:disc_formation}) the stellar and cold gas ($T_{\text{gas}} \leq 10^{4.5}$ K) components are also modelled using the same spherical multipole expansion as the halo. After disc formation, this component is instead modelled in cylindrical coordinates as a sum of Fourier terms in the azimuthal angle ($\phi$) with $m_\text{max} = 4$. The radial and vertical dependence of the coefficient of each Fourier term are interpolated (again using quintic splines) on a two-dimensional $(R, z)$ grid with 20 nodes in each direction. The resulting potential defining the disc is presented as:
\begin{equation}\label{eqn:phi_disc}
    \Phi_{\text{disc}}(R, \phi, z) = \sum^{m_\text{max}}_{m=0} \phi_{m}(R, z) e^{im\phi}.
\end{equation}
We assess the accuracy of our modelled potentials by comparing them to the stored FIRE-2 values. Because the FIRE‑2 gravitational potentials are arbitrarily normalised \citep{Wetzel_2023}, their absolute zero point cannot be compared directly to the potentials recovered by our model. To enable a meaningful comparison of the shape of the potentials, we compute the mean potential offset between our fitted potential and the FIRE‑2 values using all particles within 1$R_{\text{vir}}$. This constant offset is applied only to align the two curves for visual and structural comparison; it is not applied to our recovered potentials at any stage of the scientific analysis. In this comparison we find deviations exceeding 5\% mainly in early snapshots ($\lesssim$ 5 Gyr), when ongoing mergers perturb the system away from axisymmetry. In some simulations (e.g. m12b, m12f), later merger activity can also drive temporary excursions above the 5\% level, though these episodes are generally short-lived and rarely exceed 10\%. Overall, the {\sc agama} potentials closely follow the FIRE-2 values to within 5\% out to $\sim$50 kpc (once the galaxy has grown to at least this size). At later times ($\gtrsim$ 7 Gyr), as the system expands, this level of agreement typically extends to radii approaching $\sim$100 kpc. Since most \textit{in‑situ} and fully accreted GCs lie within these radii, we consider these deviations negligible for our analysis.

\subsubsection{Disc formation}\label{subsubsec:disc_formation}
The timing of disc formation in the simulations influences how the gravitational potentials are modelled (Section~\ref{subsubsec:methodling_potentials}), making it important to establish a clear boundary for when the disc forms. To do this, we adopt the method introduced in \citet{Correa_2017} and use the $\kappa_\text{co}$ parameter. This value quantifies the fraction of kinetic energy invested in ordered corotation and is defined as:
\begin{equation}
    \kappa_\text{co} = \frac{K_\text{rot}}{K} = \frac{1}{K} \sum_{i, L_{z,i} > 0} \frac{1}{2} m_i \left( \frac{L_{z,i}}{m_i R_i} \right)^2,
\end{equation}
where $K = 0.5 \sum m_i v_i^2$ represents the total kinetic energy of the particles and $K_\text{rot}$ is restricted to particles with positive $L_z$. Within the summation, $R_i = (r_i^2 - z_i^2)^{1/2}$ denotes the distance of the $i$th particle from the galaxy’s angular momentum axis, and $L_{z,i}$ is the component of the particle’s angular momentum along that axis.

\begin{figure}
\includegraphics[width=0.95\columnwidth]{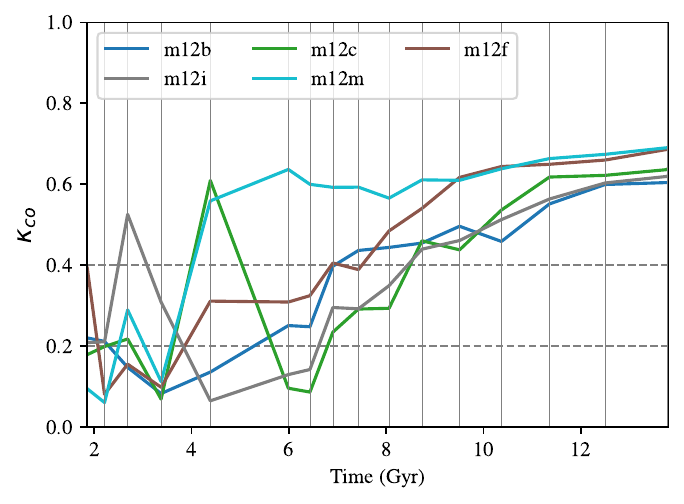}
\caption{The evolution of $\kappa_{\text{co}}$ for each simulated galaxy over time. Early spikes followed by quick declines indicate merger-induced perturbations, while subsequent steady growth marks disc formation. Each coloured line corresponds to a different galaxy, and the vertical grey lines show the snapshots where particle data is available and $\kappa_{\text{co}}$ can be calculated. Horizontal dashed lines indicate $\kappa_{\text{co}} = 0.2$, which we define as the onset of disc formation, and $\kappa_{\text{co}} = 0.4$, where the disc is considered to be formed (summarised in Table~\ref{tab:kappa_co}). The $\kappa_{\text{co}}$ values are are computed using both star and gas particles within $2R_{\star, 90}$ of each galaxy centre.}
\label{fig:kappa_co}
\end{figure}

\citet{Correa_2017} examined {\sc EAGLE} galaxies at $z=0$ and demonstrated that $\kappa_\text{co}$, when measured from star particles, serves as a reliable proxy for visual morphology, with disc galaxies typically exhibiting values greater than 0.4. It was subsequently identified by \citet{Thob_2019} that $\kappa_\text{co}$ correlates well with several internal kinematic properties, including $v_{\phi}/\sigma$, the disc-to-total mass fraction, the spin parameter, and the median orbital circularity. Extending this framework to the gas component, \citet{Jiminez_2023} found that a more stringent threshold of $\kappa_\text{co} > 0.7$ effectively selects thin, rotation-supported gas discs. In this work, we calculate $\kappa_\text{co}$ at each snapshot using both star and gas particles within $2R_{\star, 90}$. 

Following the threshold used by \citet{Correa_2017}, we classify systems with $\kappa_\text{co} > 0.4$, computed from the combined stellar and gaseous components, as disc dominated. We additionally define $\kappa_\text{co} = 0.2$ as the point at which the disc begins to form kinematically. The time evolution of $\kappa_\text{co}$ for each galaxy is presented in Figure~\ref{fig:kappa_co}, with key disc formation properties listed in Table~\ref{tab:kappa_co}. Figure~\ref{fig:kappa_co} illustrates a variation in both the onset and rate of disc formation across the sample. In particular, m12m exhibits an early and rapid rise in $\kappa_\text{co}$, crossing the kinematic formation threshold $\sim$3 Gyr earlier than the other systems. In contrast, m12b, m12c, and m12i exhibit broadly similar, more gradual increases in $\kappa_\text{co}$, while m12f shows a slower disc build-up despite beginning its kinematic disc formation at a comparable time to m12m. Early, transient spikes in $\kappa_\text{co}$ followed by rapid declines (most clearly visible in m12c and m12i) coincide with merger-driven perturbations and are not interpreted as genuine disc formation episodes. Despite these differences in formation pathways, all galaxies converge to comparable values of $\kappa_\text{co} \approx 0.6$ by $z=0$, indicating similar present-day levels of rotational support.

Our disc formation times differ slightly from those reported by \citet{McCluskey_2024}, with an average offset of 0.82 Gyr across the five galaxies. This discrepancy arises primarily from the differing definitions of disc formation. \citet{McCluskey_2024} define the onset of disc formation as the time when $(v_{\phi} / \sigma_{\mathrm{tot}}) > 1$, considering only stars that end up in the solar annulus ($6 < R < 10$ kpc and $|z| < 3$ kpc) at $z=0$, and evaluating $v_{\phi} / \sigma_{\mathrm{tot}}$ at the time of each star’s formation. Although our definition of disc formation differs methodologically from that of \citet{McCluskey_2024}, the difference of only $\sim$0.82 Gyr is small relative to the $\sim$13 Gyr evolutionary timescales probed in this work. We thus adopt our own definition of disc formation.

\begin{table}
\caption{Times at which each simulated galaxy reaches $\kappa_{\text{co}} = 0.2$ and $\kappa_{\text{co}} = 0.4$. The interval $\Delta t$ between these thresholds is defined as the kinematic disc-building time. We also present the times when ($v_{\phi} / \sigma_{\mathrm{tot}})_{\text{form}} > 1$, which \citet{McCluskey_2024} defines as the time of onset of disc formation. All values are reported in Gyr.}
\label{tab:kappa_co}
\centering
\begin{tabular}{ccccc}
\hline
Name & $t_{\kappa_{\text{co}}\text{ = }0.2}$   & $t_{\kappa_{\text{co}}\text{ = }0.4}$  & $\Delta t$  & $t_{v_{\phi} / \sigma_{\mathrm{tot}}}$\\ \hline
m12b       & 5.29  & 6.94 & 1.65 & 6.38 \\
m12c       & 6.80  & 8.49 & 1.69 & 7.31 \\
m12f       & 3.86  & 6.87 & 3.01 & 6.38 \\
m12i       & 6.61  & 8.44 & 1.83 & 7.15 \\
m12m       & 3.57  & 4.03 & 0.46 & 4.59 \\ \hline
\end{tabular}
\end{table}

\subsubsection{Orbital actions}\label{subsubsec:orbital_actions}
Under the assumption of an adiabatic and axisymmetric potential, a set of conserved quantities known as orbital actions can be used to characterise the dynamical properties of stellar orbits \citep{Binney_2008}. These actions are particularly valuable for clustering in chemo-dynamical phase space, which has become a common method in searching for stellar substructure with a shared origin \citep[e.g.][]{Myeong_2018}.

The orbital actions of the GCs are computed using {\sc agama}. This software employs the St\"ackel approximation \citep{Binney_2012}, which assumes that motion is approximately integrable and that the potential can be locally represented in a St\"ackel form \citep{de_Zeeuw_1985}. This approximation produces smooth actions because in a separable potential the motion along each coordinate evolves continuously and the actions, defined as integrals over these motions, vary gradually with small changes in position or velocity. For closed orbits, actions are defined by:
\begin{equation}
    J_i = \frac{1}{2 \pi} \oint \frac{p_{i}}{m} di,
\end{equation}
where $i \in \{R, \phi, z\}$. The radial action ($J_R$) describes how much an orbit varies in its radius and is closely related to eccentricity. The azimuthal action ($J_\phi$) is equivalent to the $z$-component of angular momentum and reflects the degree of rotational motion around the galactic centre. Finally, the vertical action ($J_z$) quantifies motion perpendicular to the galactic plane, indicating the degree to which an orbit deviates from the disc. 

While actions are conserved under the assumption of a slowly evolving potential, this condition will not hold true throughout a galaxy’s full history, particularly during phases of rapid evolution such as mergers, bar formation, or disc heating \citep[e.g.][]{Carr_2022, Pagnini_2023, Arunima_2025, Dillamore_2025, Woudenberg_2025}. Consequently, orbital actions can vary over time, limiting their reliability as tracers of dynamical history in non-steady-state systems. 

Despite this non-conservation, simulations such as those by \citet{Arunima_2025} have demonstrated that stars that formed together tend to remain correlated and move together in action space. This behaviour arises because the structures that drive changes in the potential act on physical scales much larger than the separations between stars that are born together. As a result, while one-to-one mappings between initial and final orbital states are lost, region-to-region correlations can persist and clustering in action space may still reveal information about a shared origin.

By extension, if GCs are accreted together and retain some degree of coherence in action space, clustering may reveal a shared origin between them, even after significant dynamical evolution. However, the detailed memory of their accretion is likely to be distorted, with much of the information about their progenitor systems lost. It is important to note that the simulation by \citet{Arunima_2025} considers an isolated galaxy over only a 500 Myr timescale, which limits the assessment of longer-term processes such as mergers and secular heating that could further disrupt these correlations. Consequently, the clustering of GCs with a shared origin may actually be weaker in real systems than in the stellar populations explored in that study.

\section{Orbital Evolution}\label{sec:orbital_evolution}
\subsection{Kinematic mixing}\label{subsec:kinematic_mixing}

\begin{table*}
\centering
\caption{Summary of recall, precision, and F1 scores for each simulated galaxy at $z=0$. Values represent the mean over 100 iterations of the GC formation model, with uncertainties given by the standard deviation. \textit{Ex-situ} values are similarly averaged over the 100 iterations, calculated as the mean over all \textit{ex-situ} groups in each simulation, weighted by the number of GCs in each group.}
\label{tab:mixing}
\begin{tabular}{ccccccc}
\toprule
\multirow{2}{*}{Name} & \multicolumn{3}{c}{In-Situ} & \multicolumn{3}{c}{Ex-Situ} \\
\cmidrule(lr){2-4} \cmidrule(lr){5-7}
 & Recall & Precision & F1 & Recall & Precision & F1 \\
\midrule
m12b & $0.65 \pm 0.36$ & $0.46 \pm 0.25$ & $0.51 \pm 0.25$ & $0.42 \pm 0.25$ & $0.44 \pm 0.12$ & $0.37 \pm 0.16$ \\
m12c & $0.99 \pm 0.02$ & $0.76 \pm 0.09$ & $0.86 \pm 0.06$ & $0.06 \pm 0.07$ & $0.14 \pm 0.18$ & $0.07 \pm 0.09$ \\
m12f & $0.97 \pm 0.10$ & $0.66 \pm 0.14$ & $0.78 \pm 0.12$ & $0.27 \pm 0.14$ & $0.27 \pm 0.13$ & $0.25 \pm 0.12$ \\
m12i & $0.95 \pm 0.06$ & $0.77 \pm 0.11$ & $0.84 \pm 0.06$ & $0.37 \pm 0.19$ & $0.33 \pm 0.16$ & $0.33 \pm 0.16$ \\
m12m & $0.99 \pm 0.03$ & $0.70 \pm 0.08$ & $0.82 \pm 0.06$ & $0.16 \pm 0.08$ & $0.19 \pm 0.09$ & $0.16 \pm 0.08$ \\
\bottomrule
\end{tabular}
\end{table*}

To investigate whether GCs from different origins possess unique kinematic signatures, we measure how much they mix across their orbital actions and their normalised energy, $\epsilon$, defined as:
\begin{equation}
    \epsilon = \frac{E}{|E_0|},
\end{equation}
where E is total energy of a GC and $E_0$ is the depth of the gravitational potential at the galaxy’s centre in the corresponding snapshot. Normalising by $|E_0|$ provides a dimensionless measure of energy, allowing consistent comparison of GCs across different snapshots as the potential evolves. 

\subsubsection{GC substructure detection}
As identified by \citet{Chen_2024_mar}, unsupervised clustering methods outperform supervised approaches in separating GC populations by their origins. Following their method, we first cluster the population into two groups, \textit{in-situ} and \textit{ex-situ} GCs, and then iteratively subdivide the \textit{ex-situ} population to identify individual accretion events. This process continues until the smaller of the two resulting groups contains fewer than five GCs. This is the limit below which we assume GC substructure to be undetectable. We test three unsupervised clustering methods: K-Means, Agglomerative Clustering, and BIRCH (Balanced Iterative Reducing and Clustering using Hierarchies), all of which operate without prior labels and group GCs based on similarities in their properties. Each of these methods is implemented though the {\sc scikit-learn} package \citep{scikit_learn}. We find that using the orbital quantities ($\epsilon$, $J_R$, $J_\phi$, $J_z$), Agglomerative Clustering produces the closest agreement between predicted and true groupings and so these are the results we present.

To quantify the degree of mixing from the clustering algorithm, we use the recall, precision, and F1 metrics, which compare the number of true positives (TP), false positives (FP), and false negatives (FN) produced by the clustering. 

\begin{itemize}
    \item \textbf{Recall} measures the completeness of the clustering and returns the fraction of objects from a true group that are correctly assigned to that group:
    \begin{equation}
    \text{Recall} = \frac{TP}{TP + FN}.
    \end{equation}
    
    \item \textbf{Precision} quantifies the purity of the clusters, which is the fraction of objects assigned to a cluster that actually belong to the corresponding true group:
    \begin{equation}
    \text{Precision} = \frac{TP}{TP + FP}.
    \end{equation}

    \item \textbf{F1 score} combines both recall and precision into a single metric by taking their harmonic mean and provides a quantitative measure of how well GCs of different origins remain separated in the clustering:
    \begin{equation}
    \text{F1} = 2 \times \frac{\text{Precision} \times \text{Recall}}{\text{Precision} + \text{Recall}}.
    \end{equation}
    High F1 values indicate minimal mixing, meaning GCs from distinct origins are assigned to largely pure and complete clusters, whereas lower F1 values reflect greater mixing, with clusters containing GCs from multiple origins or missing true members.
\end{itemize}

We present the results of this mixing in Table~\ref{tab:mixing}. For m12b, both the \textit{in-situ} and \textit{ex-situ} populations are recovered with limited accuracy (F1 = 0.51 and 0.37, respectively), indicating a high degree of mixing overall. By contrast, the other galaxies show relatively strong recovery of the \textit{in-situ} population, with F1 scores above 0.75. The high recall values ($\geq$0.95) confirm that most \textit{in-situ} GCs are correctly identified, although contamination from \textit{ex-situ} GCs remains significant, with precision values ranging between 0.66 and 0.77.

\begin{figure*}
    \centering
    \includegraphics[width=1\textwidth]{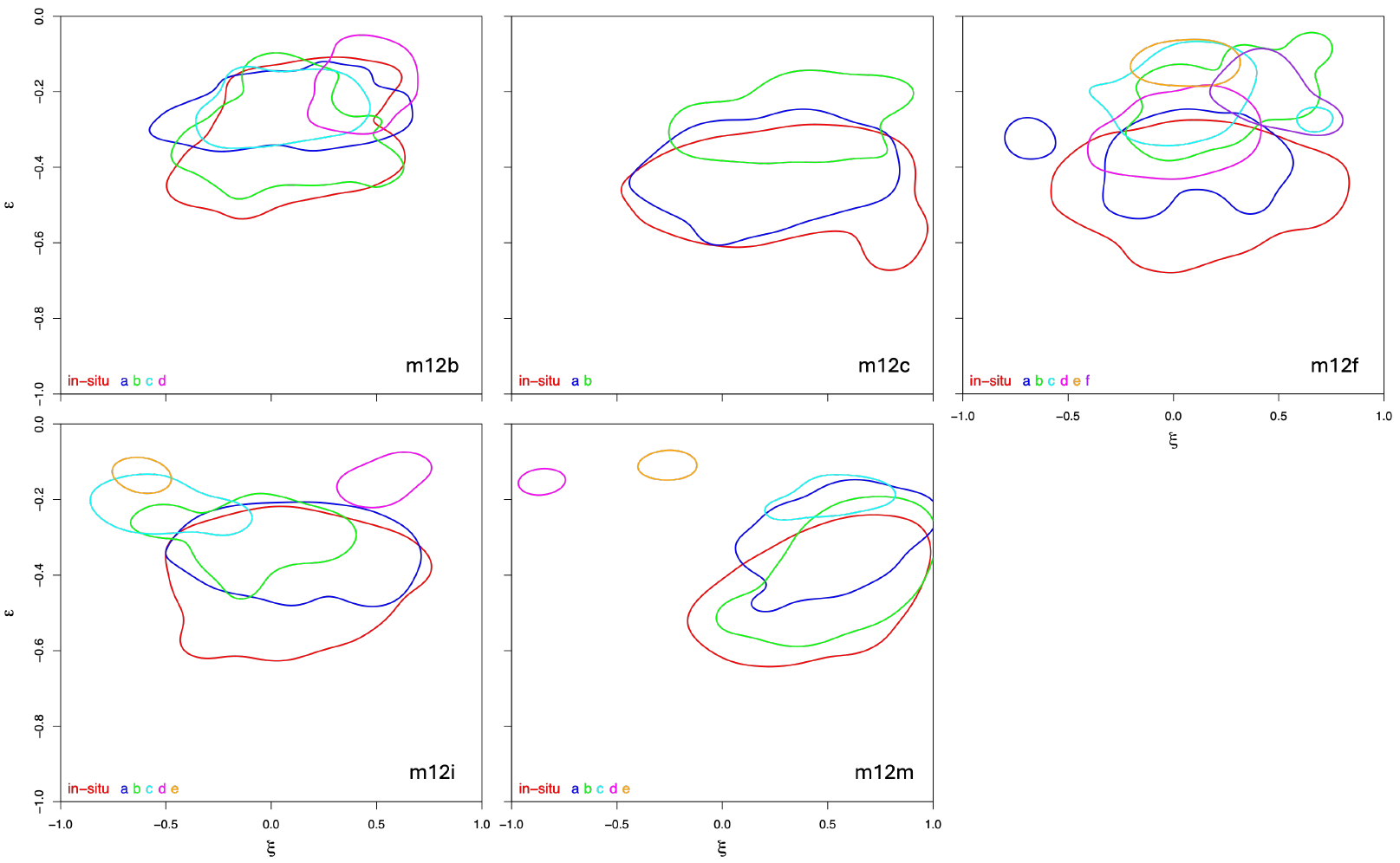}
    \caption{Normalised IoM space reflecting the normalised energy, $\epsilon = E / |E_0|$, and orbital circularity, $\xi = L_z / L_{z, \text{circ}}(E)$, for GC substructures in all simulated galaxies at $z=0$. Contours enclose the regions containing 75\% of the GCs in each group, averaged over the 100 iterations of the GC formation model. They are determined using a Gaussian KDE with bandwidths set from the variance of each variable, scaled by the square root of the sample size. Red contours show the \textit{in-situ} population, while other colours indicate distinct accretion events, ordered alphabetically by the accretion sequence in the bottom left of each panel. Large overlap across each group indicates that the different GC populations are not cleanly separated in IoM space, reflecting strong mixing between origins.
    }
    \label{fig:iom_space}
\end{figure*}

The \textit{ex-situ} GC populations are generally poorly recovered (F1 $\leq0.37$), indicating that they do not occupy distinct regions of kinematic space. Their signals are diluted both by the \textit{in-situ} population and by overlap between different \textit{ex-situ} groups. Figure~\ref{fig:iom_space} illustrates this at $z=0$, where overlapping coloured areas show GC groupings occupying similar regions of IoM space. This implies that even if individual accretion events initially contributed GCs with distinct kinematic signatures, these signatures appear to have been washed away either during the accretion process or through subsequent dynamical evolution (Sections~\ref{subsec:drivers_of_kinematic_evolution} and \ref{subsec:action_diffusion}). This severely restricts the information that present-day GC orbits can provide about their origins. 

\subsubsection{The impact of disc formation}\label{subsubsec:impact_of_disc_formation}
As revealed in Figure \ref{fig:iom_space}, m12m exhibits two small over-densities of accreted GCs that are distinct from the broader substructure distribution. This distinction comes from an offset in orbital circularity ($\xi$), which is defined as: 
\begin{equation}\label{eqn:circularity}
    \xi = \frac{L_z}{L_{z, \text{circ}}(E)},
\end{equation}
where $L_z$ is the $z$-component of the GC angular momentum and $L_{z, \text{circ}}(E)$ is the angular momentum of a circular orbit with the same energy $E$. The disc of m12m forms rapidly ($\sim$0.46 Gyr) and early ($\sim$3.5–4 Gyr; Table \ref{tab:kappa_co}), shortly after the peak of \textit{in-situ} GC formation at $\sim$3.1 Gyr. This efficient early disc growth gathers most \textit{in-situ} GCs onto prograde orbits as indicated by the positive values of $\xi$. Similarly, accretion events within the first 5 Gyr of the simulation (a, b, c) are also driven onto prograde orbits, highlighted by the corresponding coloured contours in Figure~\ref{fig:iom_space}. Following the disc orbital classifications of \citet{Yu_2023}, a substantial fraction of both these \textit{in-situ} and early-accreted GCs end up occupying thin ($\xi=0.8-1$) or thick ($\xi=0.2-0.8$) disc orbits by $z=0$. In contrast, later accretions (d and e) in m12m arrive on retrograde orbits and remain kinematically distinct in IoM space, with F1 scores of $0.76 \pm 0.36$ and $0.61 \pm 0.39$. Although individually detectable, these late events make up only 12\% and 5\% of the \textit{ex-situ} GC population at $z=0$ and so their contribution to the weighted average F1 score is minimal, explaining the low overall values reported in Table~\ref{tab:mixing}. Thus, while early disc formation allows late retrograde accretions to stand out in IoM space, the overall \textit{ex-situ} population of m12m remains largely mixed. 

By comparison, the remaining galaxies in our sample form their discs slower than m12m and at later times, allowing the \textit{in-situ} and early-accreted \textit{ex-situ} GCs to span a broader range of orbital orientations. This is in contrast to the strongly prograde orbits observed in m12m. For the other galaxies, this broader distribution means that later GC accretions ($\gtrsim$10~Gyr) are less distinguishable in IoM space, as they often occupy the same regions as the \textit{in-situ} or previously accreted GCs. Taken together, these results confirm that kinematic information alone is insufficient to reliably separate GCs by origin. This is in agreement with \citet{Chen_2024_mar}, who show that high-accuracy clustering benefits from combining kinematics and spatial properties with ages and metallicities.

\subsection{Drivers of kinematic evolution}\label{subsec:drivers_of_kinematic_evolution}

\begin{figure}
\includegraphics[width=\columnwidth]{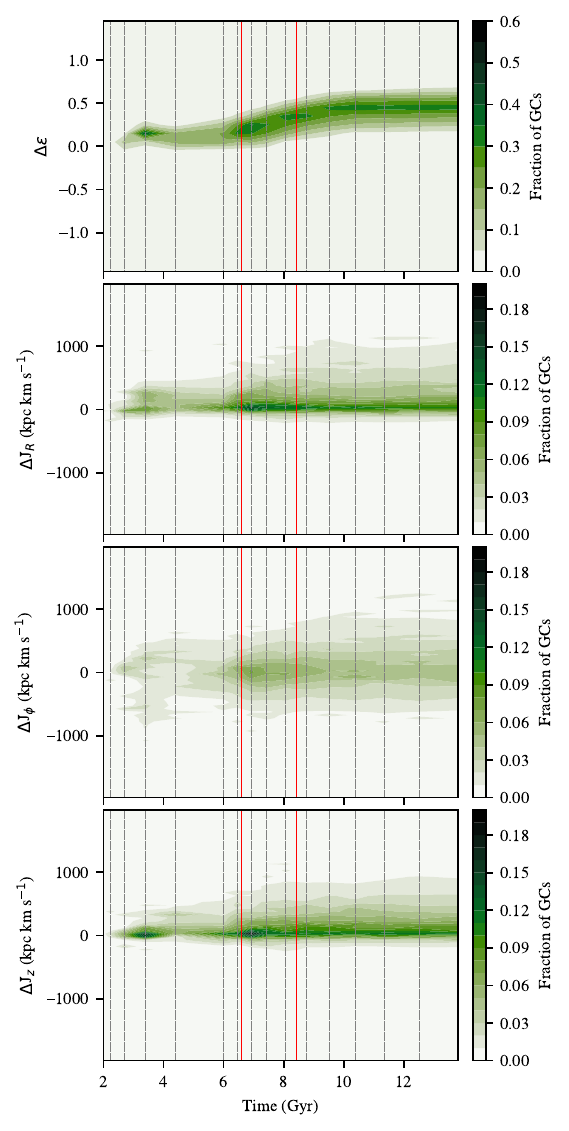}
\caption{The time evolution of kinematic changes for the \textit{in-situ} GCs in simulation m12i, including only clusters that survive to $z = 0$. No mass cut is imposed, and the surviving sample therefore spans a range of initial cluster masses and ages. At each snapshot, the cumulative change in a cluster’s kinematic quantity is obtained by summing all incremental changes from earlier times. These cumulative values are used to construct one-dimensional distributions with bin sizes of 0.1 for $\epsilon$ and 50 kpc km s$^{-1}$ for $J_R$, $J_\phi$, and $J_z$. The colour scale indicates the fraction of clusters in each bin, averaged over 100 iterations of the GC formation model. The distributions are evaluated only at the discrete simulation snapshots, which are indicated by grey vertical lines, and are displayed as a smoothly interpolated colour map to aid visual interpretation. From top to bottom, the panels present the evolution of changes in normalised energy, radial action, azimuthal action, and vertical action. Red vertical lines mark the times when $\kappa_{\text{co}} = 0.2$ and $\kappa_{\text{co}} = 0.4$, as listed in Table~\ref{tab:kappa_co}.}
\label{fig:m12i_dvar_time}
\end{figure}

Here, we investigate how GC kinematics evolve over time and the mechanisms responsible for the observed changes in kinematic quantities, using galaxy m12i as a case study. As revealed in the top panel of Figure~\ref{fig:m12i_dvar_time}, during the phase of disc growth ($\sim$6.6–8.4 Gyr), the \textit{in-situ} GC population exhibits a coherent increase in normalised energy. The panel shows this as the over-density trending upward along the energy axis, with all clusters moving in the same direction rather than scattering randomly. This illustrates what we refer to as an “ordered” change, where the population evolves coherently rather than stochastically. In contrast, the orbital actions of the GCs also evolve during this period, but their variations appear more random and diffusive, suggesting a stochastic response to perturbations. 

\subsubsection{Disc growth}\label{subsubsec:disc_growth}
Initially, the ordered increase in $\epsilon$ appears consistent with the disc-building phase of a galaxy, as the interval over which $\kappa_{\rm co}$ rises from 0.2 to 0.4 (Table~\ref{tab:kappa_co}) coincides with the period of increasing $\epsilon$ (Table~\ref{tab:time_changes}) for not only m12i but also m12b and m12f.  However, this pattern is less evident in m12c and absent in m12m. In m12m the disc forms rapidly ($\sim$0.46 Gyr) and at early times ($\sim$3.5–4 Gyr), yet the changes in kinematic quantities (especially normalised energy and radial action) occur over a much more extended period ($\sim$3.4-11.4 Gyrs). This instead suggests that rather than disc growth, a different evolutionary process drives the change in $\epsilon$. 

From this point forward, our goal is therefore to determine which evolutionary dynamic mechanism is responsible for altering the kinematics of the GC population. We continue to use the ordered evolution of $\epsilon$ as a tracer of this process, as its changes occur roughly simultaneously with variations in orbital actions (Figure~\ref{fig:m12i_dvar_time}) but are easier to interpret due to their coherent, ordered behaviour compared to the more stochastic evolution of the actions.

\begin{table}
\caption{Approximate time period over which the normalised energies change ($\Delta \epsilon$) and when average GC radial migration ($\Delta r$) becomes non-significant. The $t_{\Delta \epsilon}$ ranges correspond to the minimum and maximum times, across the 100 GC model iterations, during which the normalised energies undergo significant evolution. In this work we define radial migration as changes in the semi-major axis, $(r_{\mathrm{apo}} + r_{\mathrm{per}})/2$, of an orbit. The values listed for $t_{\Delta r}$ represent the time after which radial migration becomes non-significant on average, i.e. beyond this time the mean GC migration across the 100 iterations falls below 500 pc. All values are reported in Gyr.}
\label{tab:time_changes}
\centering
\begin{tabular}{ccc}
\hline
Name       & $t_{\Delta \epsilon}$ & $t_{\Delta r}$ \\ \hline
m12b       & 2.7 - 6.0  &  4.3   \\
m12c       & 4.4 - 10.4 &  5.8   \\
m12f       & 2.7 - 6.4  &  2.7   \\
m12i       & 3.4 - 9.5  &  3.8   \\
m12m       & 3.4 - 11.4 &  5.6   \\ \hline
\end{tabular}
\end{table}

\subsubsection{Mergers and non-axisymmetric structures}\label{subsubsec:mergers_non_axisymmetric_structures}
We next examine merger activity as a potential driver for the kinematic evolution of the \textit{in-situ} GC population but find no clear connection between merger events and the period of increasing normalised energy. Similarly, we examine the influence of massive non-axisymmetric structures, such as bars and spiral arms, traced via the $m = 2$ Fourier mode. The influence of bars and spiral structures on orbital dynamics has been well established, with resonances capable of significantly altering orbital actions. These include the co-rotation resonance \citep[CR; e.g.][]{Sellwood_2002, Roskar_2012, Daniel_2015} and the inner and outer Lindblad resonances \citep[ILR/OLR; e.g.][]{Barbanis_1967, Lynden_Bell_1972, Carlberg_1985}. Of the five galaxies studied, four (m12b, m12c, m12f, m12m) host a bar, but the epochs of peak bar strength occur around $\sim$13 Gyr \citep{Ansar_2025} which is 2–7 Gyr (depending on the galaxy) after the observed energy changes. Bars therefore seem unlikely to drive the early evolution we observe. All systems nevertheless exhibit spiral structure. Independent measurements \citep{Quinn_2025} of the Latte galaxies show that these spirals are generally weak and averaging over the five galaxies, the median of the maximum $m = 2$ Fourier amplitudes is $(A_2 / A_0)_\mathrm{max} = 0.074$. For m12b, m12f, and m12i, the earliest spiral episodes begin 0.3–3.3 Gyr after the principal changes in $\epsilon$,  indicating no temporal overlap. In m12c and m12m, the earliest spiral episodes partially overlap with the tail end of the $\epsilon$ evolution (for periods of 1.1 and 2.35 Gyr, respectively); however, by this time the dominant evolution in $\epsilon$ has already concluded, making any causal connection unlikely. Together, these results suggest that non-axisymmetric structures do not play a major role in altering GC energies. Nonetheless, during periods when bars or spiral patterns reach their maximum strength, they may still influence orbits and contribute to the non-conservation of orbital actions. Previous work by \citet{Dillamore_2025} has provided evidence that strong bars can disrupt kinematic substructure for low-energy, prograde, eccentric, or low-inclination orbits. It is therefore plausible that similar effects could act on GCs at later times, when strong non-axisymmetric structures form.

\begin{table*}
\centering
\caption{Pearson (r$_p$) and Spearman (r$_s$) correlation coefficients and corresponding p-values for the \textit{in-situ} GCs, comparing the average tidal field strength ($\overline{\Omega}_{\text{tid}}$), average normalised energy ($\overline{\epsilon}$), and central (r$_{\rm 3D} < 3$ kpc) stellar density ($\rho_{\star}$) for each simulated galaxy. Values are calculated over the period of normalised energy increase.}
\label{tab:environment_correlations}
\begin{tabular}{ccccccccccccc}
\toprule
\multirow{2}{*}{Name}
    & \multicolumn{4}{c}{$\overline{\Omega}_{\text{tid}}$ - $\overline{\epsilon}$}  
    & \multicolumn{4}{c}{$\overline{\Omega}_{\text{tid}}$ - $\rho_{\star}$} 
    & \multicolumn{4}{c}{$\overline{\epsilon}$ - $\rho_{\star}$} \\
\cmidrule(lr){2-5} \cmidrule(lr){6-9} \cmidrule(lr){10-13}
 & r$_p$ & p$_p$ & r$_s$ & p$_s$ & r$_p$ & p$_p$ & r$_s$ & p$_s$ & r$_p$ & p$_p$ & r$_s$ & p$_s$\\
\midrule
m12b & 0.67 & 5.93e-03 & 0.51 & 4.98e-02 & 0.90 & 4.03e-06 & 0.69 & 4.77e-03 & 0.78 & 5.98e-04 & 0.22 & 4.20e-01 \\
m12c & 0.59 & 3.42e-02 & 0.78 & 1.65e-03 & 0.88 & 7.95e-05 & 0.86 & 1.47e-04 & 0.89 & 4.01e-05 & 0.95 & 6.36e-07 \\
m12f & 0.70 & 3.89e-03 & 0.44 & 9.83e-02 & 0.95 & 3.85e-08 & 0.97 & 7.14e-10 & 0.82 & 2.01e-04 & 0.46 & 8.13e-02 \\
m12i & 0.96 & 2.26e-07 & 0.98 & 1.61e-09 & 0.99 & 9.14e-11 & 0.98 & 1.61e-09 & 0.96 & 1.03e-07 & 0.98 & 1.61e-09 \\
m12m & 0.81 & 4.44e-04 & 0.85 & 1.16e-04 & 0.93 & 1.86e-06 & 0.86 & 6.85e-05 & 0.94 & 6.29e-07 & 0.99 & 7.38e-11 \\
\bottomrule
\end{tabular}
\end{table*}

\subsubsection{Time-dependent potential fluctuations}\label{subsubsec:time_dependent_potential_fluctuations}
We further consider whether fluctuations in the galactic potential could be responsible for the observed evolution in $\epsilon$, and therefore for the broader kinematic transformation of the GCS, through radial migration. In lower-mass galaxies ($M_{\text{star}}(z=0) = 2 \times 10^6 - 5 \times 10^{10} M_{\sun}$), potential fluctuations from recurrent gas inflows and outflows associated with star formation can strongly perturb stellar orbits, producing alternating phases of expansion and contraction that gradually heat the stellar population \citep{El-Badry_2016}. In a more massive system like m12i, stellar feedback still generates gas flows, but the deep potential well suppresses coherent stellar velocity perturbations, limiting radial migration. Despite this suppression, we test whether the GC population undergoes any significant radial redistribution during the phase when ordered changes in their normalised energy occur. Because the GC orbits are not confined to the disc, instead of defining radial migration as changes in the guiding radius, we define it as changes in the semi-major axis ($r_{\mathrm{apo}} + r_{\mathrm{per}}) / 2$) of the orbit. 

Across all galaxies, most radial migration occurs at early times (by $\sim$5.8 Gyr at the latest), while increases in normalised energy persist to later epochs, extending to 9–11 Gyr in some systems. In most galaxies, the late stages of radial migration partially overlap with the onset of energy growth, as evidenced by the overlap between the $\Delta r$ and $\Delta \epsilon$ ranges in Table~\ref{tab:time_changes}. We interpret this as a transition in the dominant mechanism shaping GC orbital evolution. The early phase of migration reflects large-scale rearrangement of the stellar and gaseous components as the gravitational potential deepens. At these stages, the galaxies with our sample have relatively low masses (compared to those at $z=0$) and are therefore more susceptible to merger-driven perturbations, which can further contribute to the spatial redistribution of GCs. In the subsequent phase, clusters exhibit little additional radial migration and instead the increase in $\epsilon$ manifests through GC orbits becoming more eccentric and/or vertically extended. This is highlighted by the concurrent changes in the orbital actions (bottom three panels of Figure~~\ref{fig:m12i_dvar_time}) and the rise in $\epsilon$ (top panel of Figure~~\ref{fig:m12i_dvar_time}).

\subsubsection{Central mass build-up}\label{subsubsec:central_mass_build_up}
Subsequently, we find that the increase in normalised GC energy is strongly (p < 0.05) and linearly correlated with the build-up of the central stellar density ($\rho_\star$) across all galaxies in our sample. This relationship is quantified in Table~\ref{tab:environment_correlations}, where the $\overline{\epsilon}$ – $\rho_\star$ correlation yields Pearson coefficients of $r_p \geq 0.78$ for all systems, suggesting that the growth of the central potential may be the primary driver of the observed evolution in GC kinematics. Table~\ref{tab:environment_correlations} summarises the correlation coefficients and associated p-values for the \textit{in-situ} GC population for all galaxies, comparing the changes in the average tidal field strength, average normalised energy and central stellar density. All quantities are evaluated over the period during which the normalised energy increases, ensuring that the measured correlations trace the phase of strongest dynamical evolution. The reported coefficients therefore quantify the degree to which variations in the tidal environment and central stellar density are linked to the energetic evolution of \textit{in-situ} clusters. 

We measure the central density within $r < 3$ kpc of the galaxy centre where newly formed GCs are placed in the model (Section~\ref{subsubsec:gc_model}). We choose this 3 kpc aperture as it provides a tracer of the changing central gravitational potential rather than a measurement of the local density at individual GC locations. Although \textit{in-situ} GCs migrate outward, their radial distribution peaks inside 10 kpc, with most of the population contained within $\sim$20 kpc, keeping them in the region where the tidal field is shaped by the deepening central mass distribution \citep[e.g.][]{Garrison_Kimmel_2017}. During the epochs of interest, the stellar component of the galaxies remains strongly centrally concentrated, with the stellar half-mass radius generally below 3 kpc in most systems and around 5 kpc in one case (m12m), so $\rho_\star$ measured within 3 kpc continues to trace the growth of the central stellar mass that sets the inner gravitational potential. Consequently, a correlation between the tidal field experienced by a cluster and $\rho_\star$ does not rely on the clusters residing within 3 kpc at late times, as both properties track the global structural evolution of the galaxy.

Physically, the dependence of $\epsilon$ on $\rho_\star$, is expected to be mediated by the strength of the local tidal field, parametrised using the angular frequency ($\Omega_{\text{tid}}$; Section \ref{subsubsec:gc_model}). As the central stellar density increases, the gravitational potential deepens, strengthening the tidal field and facilitating the subsequent increase in GC normalised energies. However, as presented in Table~\ref{tab:environment_correlations}, the Pearson correlation between $\Omega_{\text{tid}}$ and $\epsilon$ is the weakest (0.59–0.96), compared to the stronger linear correlations between $\Omega_{\text{tid}}$ and $\rho_\star$ (0.88–0.99) and between $\epsilon$ and $\rho_\star$ (0.78–0.96). This may reflect the difference in the timescales of these quantities. Stellar density and potential depth evolve gradually over time and this evolution is driven by processes such as gas inflows, star formation and secular growth. In contrast, the strength of the tidal field reflects a more rapid response to local perturbations and asymmetries as encoded in its construction (Section~\ref{subsubsec:gc_model}). These localised variations can add noise to $\Omega_{\text{tid}}$, especially when averaged across the entire \textit{in-situ} GC population. This can weaken the correlation of $\Omega_{\text{tid}}$ with the smooth, long-term increase of $\epsilon$, even though the overall trend of the tidal field still reflects the deepening of the potential. Consequently, the evolution of $\epsilon$ is more tightly coupled to the slower structural evolution traced by $\rho_\star$ rather than to the more rapidly fluctuating $\Omega_{\text{tid}}$. Short-timescale tidal variations still induce local fluctuations in energy, consistent with the small, transient responses seen at later times (see Section~\ref{subsec:action_diffusion}), but they contribute little to this period of increasing $\epsilon$.

We also note that differences in the evolutionary pathways of each galaxy can lead to a decoupling between the strength of the tidal tensor and the change in normalised energy and an alteration of the otherwise approximately linear correlation between them. This is observed in the weaker (r$_p$ < 0.8) $\Omega_{\text{tid}}$ - $\epsilon$ correlations of m12b, m12c and m12f. To investigate this, we examine a collection of indicators tracing evolutionary history, including the specific star formation rate, cold gas fraction, degree of asymmetric structure, radial GC distribution, thin and thick disc formation \citep[with the disc components defined following][]{Yu_2023}, stellar age distribution, $\kappa_{\text{co}}$, merger history, accretion rate, and the average tidal tensor strength. We find that no single diagnostic provides a clear, universal one-to-one explanation for the inter-galaxy differences, although our ability to trace such features in detail may be limited by the temporal resolution of the available snapshots. We also search for Spearman correlations as a strictly linear correlation could obscure a monotonic but non-linear relationships between $\Omega_{\text{tid}}$ and $\epsilon$. 

In this investigation, we find that both m12i and m12m exhibit high ($r > 0.8$) Pearson and Spearman coefficients, indicating a strong linear relationship between $\Omega_{\text{tid}}$ and $\epsilon$. In m12b and m12f, $r_p > r_s$ while $r_p > 0.65$, suggesting a moderately strong and mostly linear relationship. In these galaxies, a slight peak in specific accretion rate (Appendix~\ref{asec:specific_accretion_rate}) occurs shortly after the rise in $\epsilon$, likely causes a temporary decoupling between $\epsilon$ and $\rho_\star$. This can reduce the linear correlation and also manifest as a rank inversion that suppresses the Spearman statistic. Finally, m12c shows a case where $r_s > r_p$ and $r_s = 0.78$. We attribute this to an extended period between $\sim$5–8 Gyr with minimal mass growth (Figure~\ref{fig:mass_growth}). This plateau in m12c’s mass likely delays the deepening of its potential and the corresponding increase in $\epsilon$, producing a non-linear but still monotonic relationship between $\Omega_{\text{tid}}$ and $\epsilon$.

Overall the links between $\epsilon, \Omega_{\text{tid}}$ and $\rho_\star$ align with previous work \citep{Kruijssen_2015, Pfeffer_2018}, which found that tidal heating is highly environment dependent and most effective in dense, star-forming regions. Across the GC populations, tidal heating can be localised and act non-uniformly, producing changes in radial, azimuthal, and vertical actions that reflect modifications to orbital eccentricity, inclination, and orientation. These changes can occur in a largely unordered manner, introducing stochasticity into the evolution of cluster orbits as seen in the bottom 3 panels of Figure~\ref{fig:m12i_dvar_time}, over the same period in which $\epsilon$ increases.

Although Figure~\ref{fig:m12i_dvar_time} focuses on the \textit{in-situ} population of m12i, similar trends are seen for early-accreted GCs. For mergers that occur before significant mass growth ($\sim$7 Gyr), the accreted clusters penetrate into the inner regions of the galaxy and mix with the \textit{in-situ} population, as indicated by the significant overlap of early-accreted and \textit{in-situ} contours that remain until $z=0$ (Figure~\ref{fig:iom_space}). Once occupying the same physical regions, they experience comparable tidal forces driven by the increasing stellar density. Consequently, during this phase they undergo the same ordered changes in normalised energy as the \textit{in-situ} GCs (Figure~\ref{afig:m12i_dvar_accretion}). The process of accretion however, appears to induce substantial changes in the orbital actions with a more diffuse spread as compared to the \textit{in-situ} population. In contrast, GCs accreted after the stellar density has built up show little variation in normalised energy. By this stage, the inner potential has already deepened and stabilised, so the clusters experience a relatively steady gravitational field and evolve more adiabatically. Their normalised energies remain roughly constant, however, the initial strong perturbation of the merger scatters them across a wide range of orbital actions, and their subsequent orbits largely preserve this broad dynamical spread (Figure~\ref{afig:m12i_dvar_accretion}). 

\subsubsection{A note on cluster destruction}\label{subsubsec:cluster_destruction}
When discussing tidal heating in the previous section, it is important to note that we have only examined the population of GCs that survive to $z=0$. We do not impose any explicit mass cut with the surviving sample containing a mix of initial cluster masses and ages. Consequently, the trends revealed in Figure \ref{fig:m12i_dvar_time} are subject to survivor bias, as only clusters massive enough to endure the tidal field along their orbits remain in the sample. The same tidal perturbations that drive the variations in orbital energy and actions can also accelerate structural evolution within the clusters themselves. In dense regions or environments with rapid variations in the tidal field, these fluctuations can enhance mass loss and, for sufficiently low-mass systems can lead to full cluster disruption \citep[e.g.][]{Gnedin_1997, Li_2019, Meng_2022}. A non-negligible fraction of clusters formed in the simulation are therefore destroyed before $z=0$, contributing their stars to the diffuse stellar halo. Studies of the MW GC system similarly indicate that disrupted clusters can supply a measurable fraction of halo stars and that many clusters are expected to dissolve over a Hubble time \citep[][]{Martell_2010, Baumgardt_2019, Koch_2019}. The orbital trends we report should thus be interpreted as those of the surviving subset, with low-initial mass clusters preferentially removed from the population. The question of cluster survival under different conditions is not addressed here and is left for future work.

\section{Information loss and observational implications}\label{sec:information_loss}

\subsection{The deterministic and stochastic evolution of GC orbits}\label{subsec:action_diffusion}

\begin{figure*}
    \centering
    \includegraphics[width=0.98\textwidth]{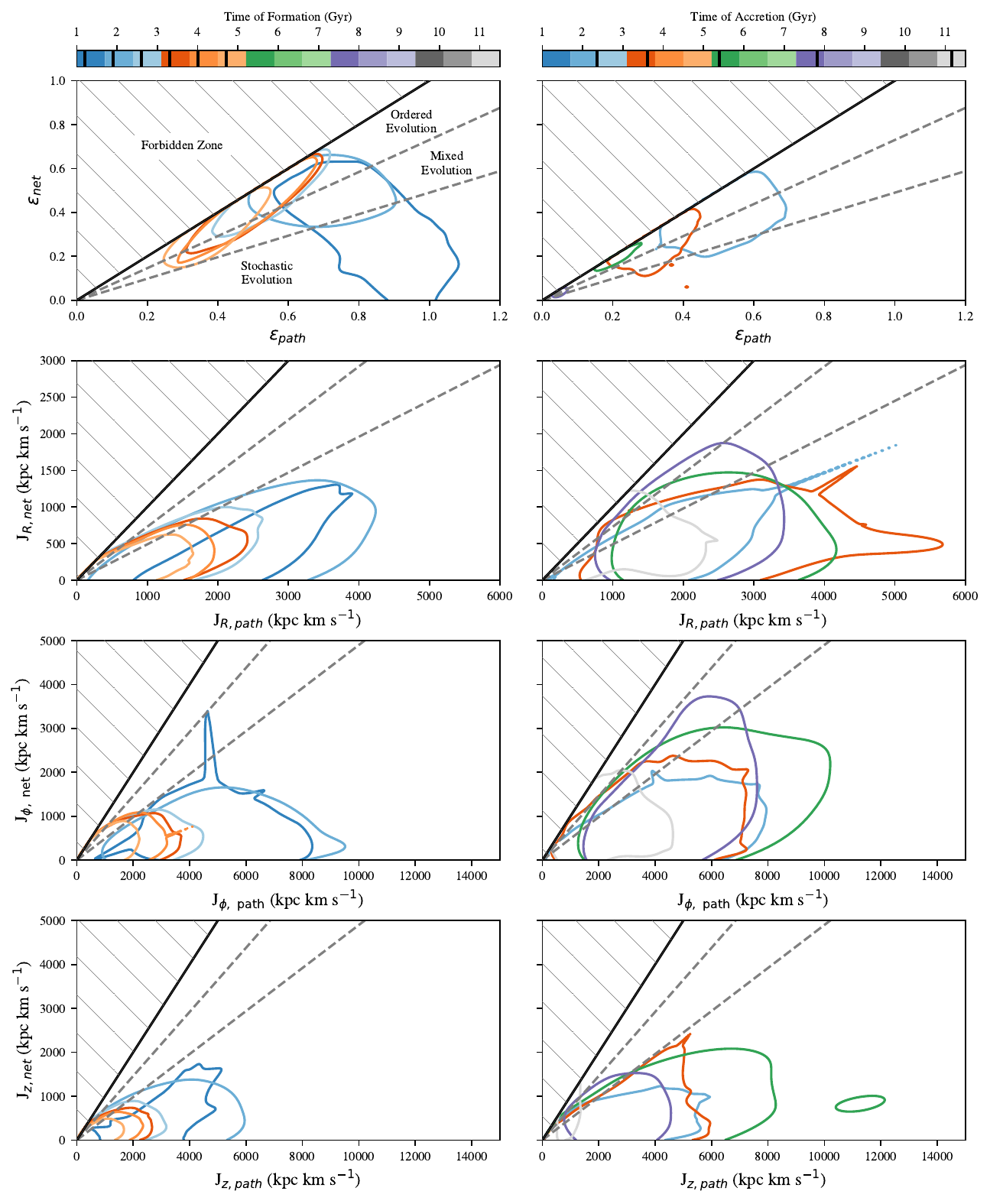}
    \caption{Comparison between total and net paths in normalised energy and orbital actions for m12i. Contours enclose 75\% of GCs in each group, averaged over 100 model iterations, determined via Gaussian KDE with bandwidths from Scott's Rule \citep{Scott_1992}. The left column shows \textit{in-situ} GCs grouped by $\sim0.7$ Gyr age bins, with paths measured from formation to $z=0$. The right column shows \textit{ex-situ} GCs grouped by accretion event, with paths measured from accretion to $z=0$. From top to bottom the rows represents normalised energy, radial, azimuthal, and vertical actions. The hatched area above the solid line marks the non-physical forbidden region (net > path). The region between the upper dashed and solid lines corresponds to ordered evolution, the lower region (below the lower dashed line) to stochastic evolution, and the middle region to mixed behaviour. For visual clarity we have labelled regions in the top left panel. Boundary definitions are detailed in Appendix~\ref{asec:random_walk_divisions}.}
    \label{fig:m12i_net_path}
\end{figure*}

As presented in Section~\ref{subsec:drivers_of_kinematic_evolution}, we have tracked the evolution of GC IoMs as the host galaxy evolves. We now quantify this evolution using two complementary measures; the cumulative walk through kinematic parameter space and the net change in each kinematic property. The cumulative walk captures the full path taken, while the net change reflects only the final displacement. For \textit{in-situ} GCs, we track this evolution from their time of formation and, for \textit{ex-situ} GCs, we begin at the point of accretion. The cumulative path length is calculated as the sum of the absolute differences in a given kinematic quantity between successive snapshots:
\begin{equation}\label{eqn:path}
    Q_{i, \text{path}} = \sum_{k=0}^{N-2}|Q_{k+1} - Q_{k}|,
\end{equation}
where $i \in \{\epsilon, J_R, J_\phi, Jz\}$ and N is the number of snapshots since formation (accretion) for \textit{in-situ} (\textit{ex-situ}) GCs. In comparison, the net change is defined as the difference between the first and last snapshot:
\begin{equation}\label{eqn:net}
    Q_{i, \text{net}} = |Q_{N-1} - Q_{0}|.
\end{equation}
Figure~\ref{fig:m12i_net_path} presents the final distributions of GCs in the Q$_{\text{net}}$-Q$_{\text{path}}$ parameter space for m12i, averaged over the 100 iterations of the GC formation model. \textit{In-situ} GCs have been grouped into age bins of $\sim$0.7 Gyrs whilst the \textit{ex-situ} population are grouped by individual accretion events. We divide this space into four regions. The uppermost region (above the solid line) of each panel represents a physically forbidden zone, as Q$_{\text{net}}$ > Q$_{\text{path}}$ is impossible for any walk. The second region (between the upper dashed line and the solid line) corresponds to lifetimes dominated by ordered (directional) changes. The lower region (between the lower dashed line and the x-axis) corresponds to stochastic, random-walk–like evolution. Finally, the middle region (between the dashed lines) reflects contributions from both ordered and random motion. The procedure used to define these boundaries is described in Appendix~\ref{asec:random_walk_divisions}. 

We continue to use m12i as a case study. As indicated by the coloured contours in the top panels of Figure~\ref{fig:m12i_net_path}, which occupy the upper regions of the parameter space, changes in normalised energy for both \textit{in-situ} and \textit{ex-situ} GCs are generally dominated by ordered evolution throughout their lifetimes. We attribute this to the increasing stellar density, strengthening the tidal tensor and injecting energy into GC orbits through orbital changes. An exception to this is seen in the earliest-formed \textit{in-situ} GCs (0.85–1.55 Gyr), which are indicated by the dark blue contours in the top left panel of Figure~\ref{fig:m12i_net_path} and stretch across all three regions of the space. These GCs formed during a period when the early progenitor of m12i experienced a major merger ($\sim$1:1.2). The resulting asymmetries in the potential impacted individual GCs differently depending on their positions, producing stochastic energy changes that were strong enough to dominate over the lifetime of this early \textit{in-situ} group of GCs. 

For the \textit{ex-situ} population of GCs, those which were accreted early (< 4 Gyr; blue and red contours in the right panels of Figure~\ref{fig:m12i_net_path}) show a slightly broader spread across regions in $\epsilon$, indicating a more stochastic evolution as compared to later accreted GCs which are constrained to the top region of the parameter space. This reflects energy fluctuations prior to the galaxy’s potential fully strengthening, as was seen in the early formed \textit{in-situ} GCs, although to a lesser extent. By contrast, orbital actions for both \textit{in-situ} and \textit{ex-situ} GCs are dominated by stochastic changes, with most age contours confined to the lower regions of each panel. This aligns with the results displayed in Figure~\ref{fig:m12i_dvar_time}, where it is demonstrated that orbital action changes appear to be random in nature and dispersed over a wide range of values. This result is also consistent with \citet{Arunima_2025}, who found that even in isolated MW-like disc galaxies, stellar orbital actions undergo diffusive, random-walk-like evolution. As such, in a non-adiabatically growing potential such as m12i, it is not surprising to find that stochastic changes naturally dominate GC actions.

\begin{figure}
\includegraphics[width=0.9\columnwidth]{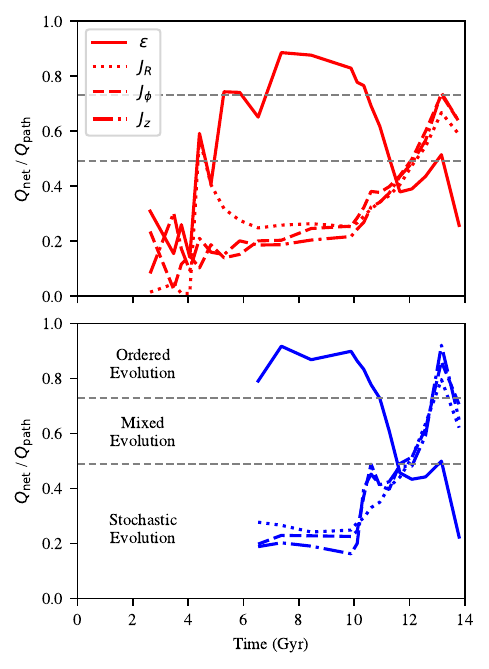}
\caption{Moving window analysis of the ratio Q$_{\text{net}}$/Q${_\text{path}}$ for m12i, averaged over 100 iterations of the GC model. Each window contains 10 snapshots over which Q$_{\text{net}}$ and Q${_\text{path}}$ are calculated. The top panel represents the \textit{in-situ} population and the bottom panel is the \textit{ex-situ}. Each panel is divided into three regions, separated by the dashed lines. The upper region corresponds to ordered evolution, the lower region to stochastic evolution and the middle region to mixed behaviour. For visual clarity we have labelled regions in the bottom panel.} The legend in the top panel also applies to the bottom panel and is independent of colour. Boundary definitions are detailed in Appendix~\ref{asec:random_walk_divisions}.
\label{fig:m12i_k}
\end{figure}

To further investigate periods where stochastic or ordered evolution dominates, we implement a moving window analysis of the ratio Q$_{\text{net}}$/Q${_\text{path}}$, averaged over all \textit{in-situ} and \textit{ex-situ} GCs, as presented in Figure~\ref{fig:m12i_k}. Due to the irregular spacing of snapshots in the first FIRE-2 public release, we use a window containing 10 snapshots rather than a fixed duration in time. We chose this window size as we find it offers temporal resolution whilst minimising noise. This window is then advanced by one snapshot at a time to trace how the relative contributions of ordered and stochastic evolution vary continuously over the galaxy’s lifetime. For each window, we compute the mean of the ratio across all GCs, allowing us to identify epochs where the dynamical behaviour of the GC system transitions between coherent, systematic evolution to more stochastic variations. As in Figure~\ref{fig:m12i_net_path}, larger values of Q$_{\text{net}}$/Q${_\text{path}}$ indicate ordered changes, while smaller values represent random-walk-like behaviour. Because the window spans only 10 snapshots, the separation between different regions of Q$_{\text{net}}$/Q${_\text{path}}$ is not sharply defined. This issue is compounded in the final 11 snapshots (13.78-13.80 Gyr), where the spacing is only 2.2 Myr, compared to an average of 400 Myr across all snapshots. The denser time spacing causes variations in the Q$_{\text{net}}$/Q${_\text{path}}$ separations relative to those used previously. Nevertheless, we retain the same boundaries as in Figure~\ref{fig:m12i_net_path}, as they provide a useful approximation (Appendix~\ref{asec:random_walk_divisions}).

Focusing on the \textit{in-situ} population of GCs, both the normalised energy and actions are initially dominated by stochastic behaviour (Figure~\ref{fig:m12i_k}; $\lesssim$5.5 Gyr). This reflects the influence of early high mass ratio mergers and their strong perturbative impact on the GC orbits present at that time. From this point on, both the \textit{in-situ} and \textit{ex-situ} GCs present very similar trends Q$_{\text{net}}$/Q${_\text{path}}$. As the inner stellar density of the galaxy increases, the potential deepens and stabilises, leading to a transition in the normalised energy toward ordered evolution ($\sim$6-10 Gyr). During this same period the actions remain dominated by stochastic changes as described in Section~\ref{subsec:drivers_of_kinematic_evolution}. After $\sim$10 Gyr, we observe a sharp drop in $Q_{\mathrm{net}} / Q_{\mathrm{path}}$ for $\epsilon$, indicating that fluctuations again begin to dominate. Examining the $\epsilon$ evolution of $Q_{\mathrm{path}}$ and $Q_{\mathrm{net}}$, after this sharp transition, we see that the path and net changes have declined to approximately 0.03 and 0.01, respectively, about 1/25 and 1/80 of their peak values during periods of ordered migration. This indicates that the system has largely stabilised. Although these variations are significantly smaller than before, the larger path relative to the net change suggests that the remaining fluctuations are driven by transient effects, such as passing substructures or localised asymmetries in the tidal field rather than by the long term evolution of the potential. These fluctuations are much smaller than earlier changes and thus do not noticeably affect the overall path over the system’s lifetime, as seen in the top panels of Figure~\ref{fig:m12i_net_path}. Regarding the orbital actions (which follow similar $Q_{\mathrm{net}} / Q_{\mathrm{path}}$ behaviour to each other), we observe a substantial increase in $Q_{\mathrm{net}} / Q_{\mathrm{path}}$, indicating that ordered motion is beginning to dominate as the galactic potential evolves toward a smoother, more adiabatic state at $\sim$10 Gyr. As described in Section~\ref{subsubsec:orbital_actions}, actions are computed using the St\"ackel approximation, which assumes that the motion of the GCs evolves continuously and the actions vary gradually with small changes in a GC's position or velocity. However, large-scale structures such as bars, spiral arms or substructures can perturb the local GC positions and velocities, resulting in deviations of the action estimates. In particular, near resonances (CR/ILR/OLR) or in regions with strongly non-integrable dynamics, the St\"ackel approximation may no longer accurately capture the true actions, leading to increased scatter. We note again that for m12i a strong bar is never fully found to form \citep{Ansar_2025}, however, m = 2 spiral episodes are seen from 9.80 Gyrs onwards \citep{Quinn_2025}. As a result, at later times ($\gtrsim$12 Gyr) in the galaxy's evolution the actions occupy an intermediate region of the $Q_{\mathrm{net}} / Q_{\mathrm{path}}$ parameter space, reflecting the combination of a nearly adiabatic potential and the residual effects of galaxy structures. 

We see similar behaviour across the other galaxies within our sample, however, we would briefly like to highlight one observed difference. As outlined in Section~\ref{subsec:drivers_of_kinematic_evolution}, galaxies m12b and m12f display a decoupling between $\epsilon$ and $\rho_\star$ following accretion events that are sufficiently large to generate local peaks in their specific accretion rates (Appendix~\ref{asec:specific_accretion_rate}). This decoupling is also reflected in the moving-window analysis of the ratio $Q_{\mathrm{net}} / Q_{\mathrm{path}}$ for each system (Appendix~\ref{asec:impact_of_mergers}). In contrast to m12i, which exhibits a relatively ordered evolution in $\epsilon$, these mergers induce stochastic variations in the kinematic evolution of m12b and m12f. The resulting fluctuations are significant enough to affect the lifetime cumulative $Q_{\mathrm{net}}$ and $Q_{\mathrm{path}}$ values, causing these to predominantly occupy regions characteristic of a mixed walk (Appendix~\ref{asec:random_walk_divisions}). This underscores the complicating influence of mergers when attempting to reconstruct a galaxy’s kinematic history. 

\subsection{Recoverability of progenitor kinematics}\label{subsec:recoverability_of_progenitor_kinematics}

\begin{figure*}
    \centering
    \includegraphics[width=0.7\textwidth]{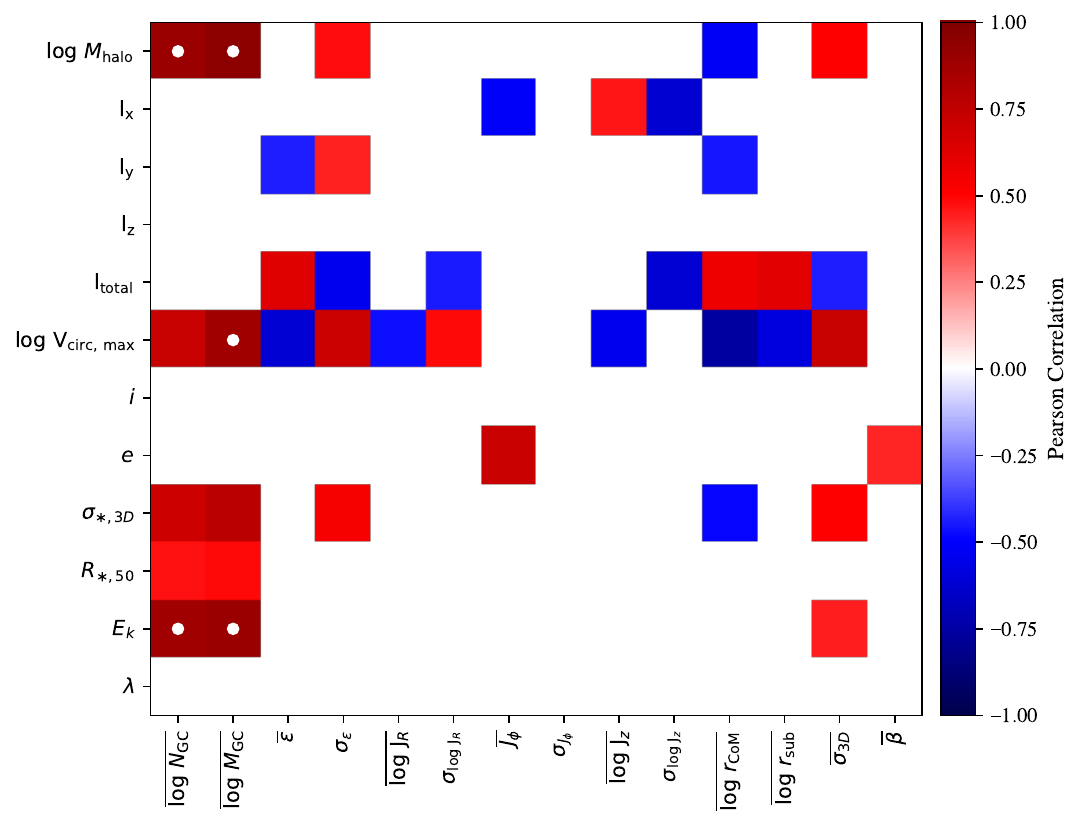} 
    \caption{Pearson correlation matrix between the average properties of GC substructure at $z=0$ (averaged over 100 iterations of the GC model) and the characteristics of their progenitor galaxies at the time they first became satellites of the host halo. These correlations have been evaluated across all five galaxies within the sample to capture general trends. We highlight strong correlations ($|r| \geq 0.8$) with white dots. White spaces indicate relationships which are not significant ($p>0.05$).}
    \label{fig:correlations}
\end{figure*}

In this work, we have demonstrated that from the time of accretion, GCs experience a range of effects that can alter their kinematic properties. This suggests that the original kinematic signal of GCs at the point of accretion becomes progressively washed out by the subsequent evolution of the host galaxy. Such evolution may occur through the merger process itself or through mechanisms such as the buildup of the central stellar density. We also identify that asymmetric structures in the galaxy can induce kinematic changes, although these appear to have a smaller impact than the previously discussed evolutionary processes. 

The time of accretion further influences the extent to which the GC signal is washed away. GCs accreted before the main build-up of the galaxy experience a more turbulent phase, during which the galaxy is still assembling its fundamental structure. These clusters undergo substantial changes in their energies that are not experienced by GCs accreted at later times (see Section~\ref{subsubsec:central_mass_build_up}). Consequently, GCs accreted from different progenitor galaxies are subject to distinct kinematic transformations. Moreover, even GCs accreted from the same galaxy and sharing a common origin can evolve differently, as demonstrated by single GC contours occupying a broad range of net and path values in Figure~\ref{fig:m12i_net_path}. This large spread in net values helps explain why the contours in the normalised IoM space of Figure~\ref{fig:iom_space} occupy such an extended region, with some GCs experiencing significantly larger changes than others within the same group. All of these effects cause accreted GC substructure to lack a distinct kinematic signal, as was highlighted by the F1 scores presented in Table~\ref{tab:mixing}.

To assess the extent to which GC substructures retain the kinematic signatures originally imprinted by their progenitor galaxies, Figure~\ref{fig:correlations} presents the Pearson correlation matrix between their mean properties at $z = 0$ and the properties of their progenitor galaxies at the time they first became satellites of the host halo (at infall). \citet{Chen_2024_mar} present a similar matrix in their work, however, we focus more on the kinematic properties of the merger from which the GC substructure was accreted. To ensure that we identify global trends, the matrix in Figure~\ref{fig:correlations} is the result of evaluating GC substructure across all galaxies in our sample. Each substructure property is averaged over 100 iterations of the GC formation model, yielding 14 quantities that describe their present-day configuration: the average number (N$_{\mathrm{GC}}$) and total mass (M$_{\mathrm{GC}}$) of GCs; the mean and standard deviation of the integrals of motion ($\epsilon, J_{R}, J_{\phi}, J_{z}$); the galactocentric distance of the substructure’s centre of mass ($r_{\text{com}}$); the mean distance of member GCs from this centre ($r_{\text{sub}}$); the three-dimensional velocity dispersion ($\sigma_{\mathrm{3D}}$); and the orbital anisotropy parameter. This is defined as:
\begin{equation} \beta = 1 - \frac{\sigma_{\theta}^2 + \sigma_{\phi}^2}{2\sigma_{r}^2}.
\end{equation}
These are compared to 12 properties of the corresponding progenitor galaxies, measured at infall: halo mass ($M_\mathrm{halo}$); specific angular momentum in both Cartesian components ($l_x, l_y, l_z$) and in total ($l_{\mathrm{total}}$); maximum circular velocity of the halo ($V_{\mathrm{circ, max}}$); orbital inclination (i) and eccentricity (e); 3D stellar velocity dispersion ($\sigma_{\star \mathrm{, 3D}}$); stellar half-mass radius (R$_{\star \mathrm{, 50}}$); kinetic energy (E$_{\mathrm{k}}$); and the Bullock spin parameter \citep[$\lambda$][]{Bullock_2001}. In the matrix presented in Figure~\ref{fig:correlations}, we highlight strong correlations (|r$_p| \geq 0.8$) with white dots. A small subset of progenitor properties show a clear correspondence with both the number and total mass of GCs. Unsurprisingly, the total GC mass correlates strongly with progenitor halo mass, a relationship well established in previous studies \citep[e.g.][]{Harris_2015, Forbes_2018_dec, Chen_2023, Chen_2024_mar}, which indicates that this relation follows a linear scaling. In our analysis, this relationship can be expressed as:
\begin{equation}
    M_{\text{GC}} \approx 1.7(\pm 0.2) \times 10^{-5} M_{\text{halo}},
\end{equation}
which is shallower than the relation reported in \citet{Chen_2024_mar}:
\begin{equation}
    M_{\text{GC}} \approx 3 \times 10^{-5} M_{\text{halo}}.
\end{equation}
This is not a large difference and may reflect variations introduced by averaging across multiple iterations of the GC model. 

We also note that the other progenitor properties that correlate with $M_{\text{GC}}$ are themselves correlated with $M_{\text{halo}}$. To isolate the dominant driver, we perform a partial correlation analysis controlling for halo mass. After doing so, all other correlations drop below |r$_p| < 0.25$, indicating that $M_{\text{halo}}$ is the primary factor underlying these trends. 

Similarly, we find that $N_{\text{GC}}$ correlates strongly with halo mass, consistent with both observational results and previous simulations \citep[e.g.][]{Choksi_2019, Burkert_2020, Zaritsky_2022, Chen_2023}. Our best-fit relation: 
\begin{equation}
    \log \text{N}_{\text{GC}} = -10.00(\pm 1.31) + 1.04 (\pm 0.12) \times \log M_{\text{halo}},
\end{equation}
agrees closely with the observational relation of:
\begin{equation}
    \log \text{N}_{\text{GC}} = -9.58(\pm 1.58) + 0.99 (\pm 0.13) \times \log M_{\text{halo}},
\end{equation}
from \citet{Burkert_2020}, for galaxies over the minimum mass limit of $M_{\text{halo}} > 10^{10} M_{\sun}$. Although $N_{\text{GC}}$ also shows apparent correlations with $E_k$, it weakens substantially (|r$_p| < 0.27$) once halo mass is controlled for, indicating that the dependencies are primarily driven by mass.

Finally, although it does not meet our threshold for a strong correlation, we find a moderately strong negative correlation (|r$_p| > 0.75$) between the present-day galactocentric distance of GC substructures and the maximum circular velocity of their progenitor galaxies at infall. Unlike the previously identified relations, the correlation with $V_{\text{circ, max}}$ (r$_p = -0.76$) is stronger than that with halo mass ($M_{\text{halo}}$; $r = -0.52$). \citet{Pfeffer_2020} has previously identified that earlier mergers and more massive progenitors tend to deposit GCs on orbits closer to the galactic centre, motivating us to test for the influence of these factors. Controlling for merger time using a partial correlation has no effect on the relationship between $r_{\text{com}}$ and $V_{\text{circ, max}}$ and when controlling for halo mass it only slightly weakens the correlation to r$_p = -0.74$. As such, there appears to be a direct link between $r_{\text{com}}$ and $V_{\text{circ, max}}$ and although it is not significant (|r$_p| < 0.80$), the correlation remains noteworthy. For a given halo mass, a higher V$_{\text{circ, max}}$ indicates a more concentrated progenitor ($V_{\text{circ}}^2 / G = M/r$) with a denser central region and deeper potential well. Because such satellites are more tightly bound, they will lose mass more slowly to tidal stripping and retain a larger effective mass during infall. Since the efficiency of dynamical friction scales with satellite mass \citep{Binney_2008}, these more compact, high-V$_{\text{circ, max}}$ galaxies experience stronger drag and sink more efficiently towards the host's centre before disruption. Their GCs are therefore deposited at smaller present-day galactocentric distances with our best-fit relation to the model data being:
\begin{equation}
    \log r_{\text{com}} = 7.54(\pm 1.26) - 3.48 (\pm 0.67) \times \log \text{V}_{\text{circ, max}}.
\end{equation}

In summary, we find that there are few strong (|r$_p| \geq 0.8$) relationships between the properties of an \textit{ex-situ} GC population at $z=0$ and those of its progenitor galaxy. The most significant correlations are between halo mass and both the total mass and number of GCs. After accounting for partial correlations, only one kinematic link remains noteworthy. This is the correlation between the average galactocentric distance of the GC substructure’s centre of mass and the progenitor’s maximum circular velocity. This supports the idea that much of the original kinematic information is erased during accretion and subsequent evolution, and that GCs retain little memory of their progenitor’s dynamics.

\section{Conclusions}\label{sec:conclusion}
In this work, we have used the GC formation model from \citet{Chen_2024_jan} coupled to MW analogs from the Latte suite of the FIRE-2 simulations \citep{Wetzel_2023} to evaluate the extent to which GC kinematics change over the evolution of a host galaxy. This has been performed with a particular focus on changes to GC normalised energy and orbital actions. 

We find that, at $z=0$ there is significant overlap of GC substructure across kinematic parameter space, such that \textit{in-situ} and \textit{ex-situ} GCs do not necessarily occupy distinct regions of this space (Table~\ref{tab:mixing}). We find that the extent to which kinematic mixing occurs is dependent upon a combination of accretion properties and host evolution. For example, as discussed in Section~\ref{subsec:kinematic_mixing}, the early and efficient disc formation in m12m results in retrograde accreted GCs being more kinematically distinct than their counterparts in galaxies where disc formation occurs later and over more extended timescales. This significant overlap across kinematic parameter spaces means that identifying GCs from different origins can be a near impossible task when using kinematic properties alone. Instead, kinematic and spatial information in combination with ages and metallicities is needed to achieve the high clustering accuracy presented in \citep[e.g.][]{Chen_2024_mar}.

To better understand why it is that kinematics alone are not appropriate in identifying distinct GC substructures, we search for evolutionary processes that may cause kinematic signals of different origins to be washed away. In this search we find that over cosmic time, \textit{in-situ} GC populations undergo a coherent increase in normalised energy and, over the same time period, experience stochastic changes in orbital actions that do not follow an obviously ordered evolution (Figure~\ref{fig:m12i_dvar_time}). By looking at the evolution of the host galaxy, we find that these kinematic changes correlate with the increase in central stellar density (Table~\ref{tab:environment_correlations}). Hence, as the host galaxy builds up its central structure, \textit{in-situ} GCs are pushed onto new orbits and the original signal of their formation disappears. The strength and shape of this correlation differs across the five galaxies analysed, which we attribute to variations in their evolutionary histories. 

For \textit{ex-situ} GCs, the time at which they are accreted impacts the extent to which their kinematics change. For GCs accreted prior to the build up of central stellar density, we note a change in normalised energy that follows a similar pattern to that of the \textit{in-situ} population. However, the orbital actions of \textit{ex-situ} GCs have more variation which appears to be mostly driven by the turbulent merger process. For GCs accreted later in a galaxy's evolution, the changes in normalised energy are less substantial, although the orbital actions still experience significant variation. We again attribute this to be driven by the merger process. Combining these observations across the \textit{in-situ} and \textit{ex-situ} GC populations, we conclude that both the deterministic and stochastic evolution of kinematic properties leads to diffuse signals of substructure at $z=0$, making the identification of origin a difficult task. Furthermore, the evolution of \textit{ex-situ} GCs largely erases the kinematic tracers of their progenitor galaxies present at infall, which would otherwise encode information about the merger process. Despite this substantial loss of information, a small number of correlations persist (Figure~\ref{fig:correlations}). In particular, the progenitor halo mass remains correlated with both the total mass and number of \textit{ex-situ} GCs at $z=0$. Beyond this, the only other surviving signal links the galactocentric distance of the GC substructure’s centre of mass to the progenitor’s maximum circular velocity.

In summary, the extent of kinematic mixing and the dynamical drivers behind it ultimately set the limits on what we can recover about GC origins by acting to erase the signatures needed to extract this information from kinematics alone. As a result, kinematic mixing rapidly weakens our ability to distinguish accreted clusters from \textit{in-situ} ones, to identify clusters that formed or were accreted as part of the same event, and to reconstruct the properties of the progenitor systems in which they originally formed.

\section*{Acknowledgements}
We thank the referee for helpful and constructive comments that improved the manuscript. We also thank Yingtian (Bill) Chen for assistance in applying the globular cluster formation model to the FIRE-2 simulations. This research made use of the Katana computational cluster, supported by Research Technology Services at UNSW Sydney. SLM and FP acknowledge support from the UNSW Scientia Fellowship programme. FP also acknowledges support from the Australian Government Research Training Program Scholarship. EJI acknowledges the support of the Australian Research Council through Discovery Project DP220103384.

\section*{Data Availability}
Simulation outputs for all of our GC iterations are available on request to the corresponding author. The public data release of the FIRE-2 Latte suite is available at \url{https://flathub.flatironinstitute.org/fire}.



\bibliographystyle{mnras}
\bibliography{references} 




\appendix

\section{Specific Accretion Rate}\label{asec:specific_accretion_rate}
\begin{figure*}
    \centering
    \includegraphics[width=1\textwidth]{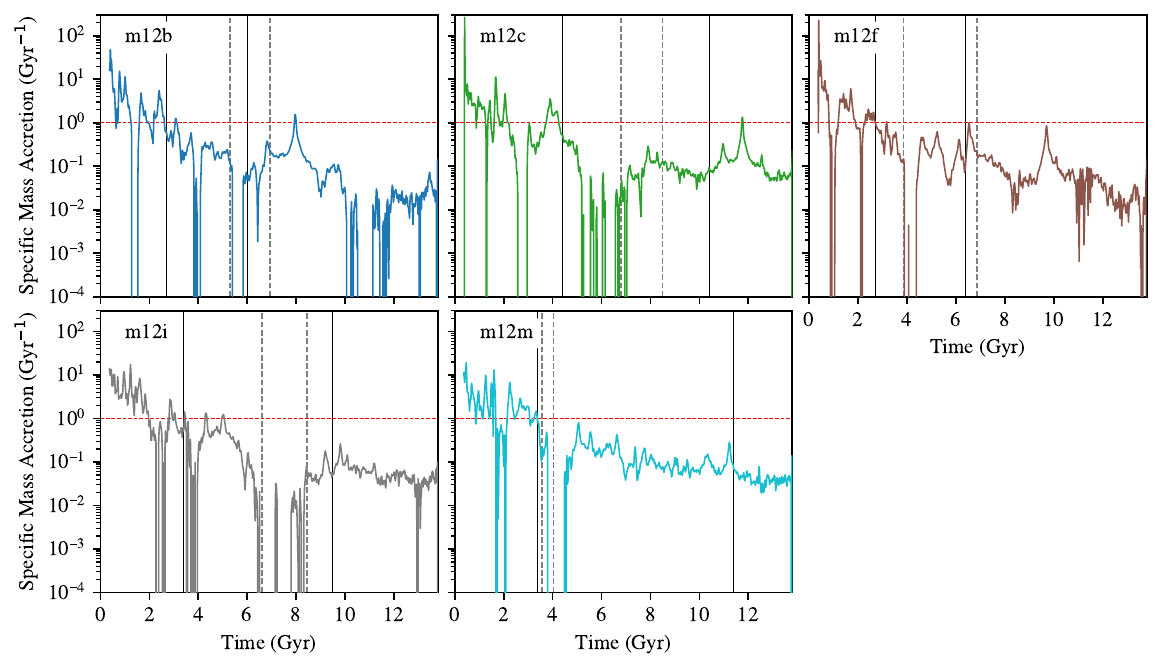}
    \caption{Specific accretion rate ($\dot{M} / M$) for each galaxy within our sample. The dotted horizontal line represents the value $p_3$ = 1 Gyr$^{-1}$ over which a GC formation event is triggered. The dashed vertical lines represent the disk building phase (see Table~\ref{tab:kappa_co}) and the solid vertical lines represent the period in which $\epsilon$ is seen to increase (see Table~\ref{tab:time_changes}). We log the y-axis for visual clarity.}
    \label{afig:specific_accretion}
\end{figure*}

In Figure~\ref{afig:specific_accretion} we present the specific accretion rate for each galaxy within our sample. We attribute the decoupling of $\epsilon$ and $\rho_\star$ in m12b and m12f to mergers significant enough to cause local peaks in specific accretion rate between $\sim$6-9 Gyrs. This is also reflected in Appendix~\ref{asec:impact_of_mergers} and is the period directly following the increases in $\epsilon$. We believe the late merger causing the peak at $\sim$12 Gyr in m12c is too late to significantly impact the $\epsilon$-$\rho_\star$ relation. 

\section{Ex-situ Kinematic Evolution}\label{asec:ex_situ_kinematic_evolution}
\begin{figure*}
\includegraphics[width=0.8\textwidth]{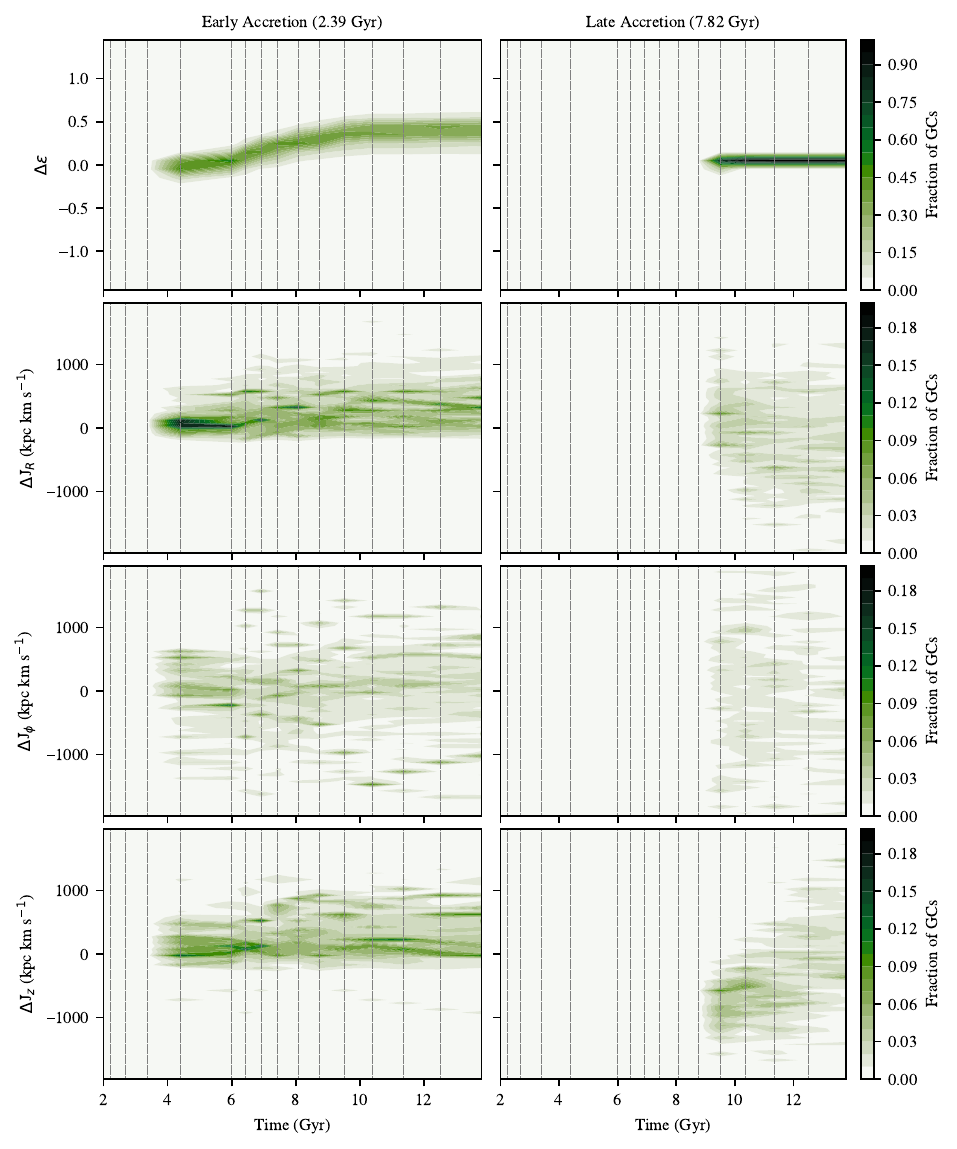}
\caption{Time evolution of kinematic changes for two accreted groups of GCs in simulation m12i, including only clusters that survive to $z = 0$. The left column represents the early accreted (2.39 Gyr) group of GCs. Listed as group a in Figure~\ref{fig:iom_space}. The right column represents the later accreted (7.82 Gyr) group of GCs. Listed as group d in Figure~\ref{fig:iom_space}. From top to bottom, each panel shows the distribution of cumulative changes in normalised energy, radial action, azimuthal action, and vertical action. Colours show the fraction of clusters within each bin, averaged over 100 iterations of the GC formation model. Vertical grey lines mark snapshot intervals.}
\label{afig:m12i_dvar_accretion}
\end{figure*}

Each panel of Figure~\ref{afig:m12i_dvar_accretion} is similar to Figure~\ref{fig:m12i_dvar_time} in the main text, but instead shows results for two groups of \textit{ex-situ} GC. The first group was accreted before (2.39 Gyr) significant mass growth ($\sim$ 7 Gyr), whilst the later group was accreted after (7.82 Gyr). For the earlier accretion, the GCs undergo the same ordered changes in normalised energy as the \textit{in-situ} GCs, however their orbital actions appear more diffuse as compared to the \textit{in-situ} GCs. We attribute this to the merger process creating local environmental changes that are non-adiabatic. This results in the assumption that the potential can be described by a St\"ackel form no longer holding true. As such, the GCs are seen to occupy a wide range of actions. For the later accretion, their normalised energies remain roughly constant, but their orbital actions continue to change and occupy a wide range of values. We attribute this to the same processes driving the orbital action spread seen in the early accretion.

\section{Random Walk Divisions}\label{asec:random_walk_divisions}
In this section, we present a simple model that defines the boundaries of Q$_{\text{net}}$ / Q$_{\text{path}}$, indicating where deterministic and stochastic migration dominate a given walk. The model simulates the motion of a set of particles along a line, with each particle’s movement consisting of a mix of predictable (deterministic) and random (stochastic) components. For the FIRE-2 public data release, the available full particle snapshots are not evenly spaced and so we adopt the same time spacing in this model to enable a direct comparison to our GC results.

\subsection{Model Setup}\label{asubsec:method_setup}
For each particle:
\begin{enumerate}
    \item Deterministic steps advance at a constant velocity drawn from a uniform distribution $v\sim U(0,1)$, with step sizes proportional to the time interval between consecutive snapshots (step = $v\Delta t$).
    \item Stochastic steps are defined by drawing a value $X \sim \mathcal{N}(0,\sigma^{2})$ with $\sigma = 0.4$, and scaling it with the square root of the time interval between consecutive snapshots ($\text{step} = X \sqrt{\Delta t}$). This produces diffusion-like behaviour and is motivated by \citet{Arunima_2025}, who demonstrated that diffusion-like random walks effectively model changes in stellar orbital actions.
\end{enumerate}

We specify the fraction of steps that are deterministic and randomly select which time intervals correspond to deterministic motion, with the remaining intervals corresponding to stochastic motion. This ensures that deterministic and stochastic steps are evenly distributed over time and are not biased toward longer or shorter intervals. To prevent either component from dominating purely because of its numerical scale, we adjusted the ranges of the uniform (deterministic) and normal (stochastic) distributions so that they produce step sizes of similar magnitude when evaluated over the same time intervals. This guarantees that any differences in behaviour arise from the relative ordering and frequency of deterministic versus stochastic steps, rather than from an arbitrary amplitude mismatch. Accordingly, we normalise the velocity and stochastic amplitudes so that their typical step sizes are similar. Because the model is concerned with the relative dominance of deterministic and stochastic motion rather than their absolute scales, this normalisation does not affect the resulting Q$_{\text{net}}$/Q$_{\text{path}}$ divisions.

\subsection{Model Procedure}\label{asubsec:method_preocedure}
We run the model for 100,000 particles, tracking their full trajectories and calculating both the net displacement and the total path length for each particle. We consider three fractions of deterministic steps (25\%, 50\%, and 75\%) to study how the distribution of Q$_{\text{net}}$ / Q$_{\text{path}}$ varies with the mix of motion types. Particles below the 25–50\% boundary are classified as dominated by stochastic motion, while those above the 50–75\% boundary are dominated by deterministic motion. The intermediate zone represents a mixture of the two behaviours.

\begin{figure}
\includegraphics[width=\columnwidth]{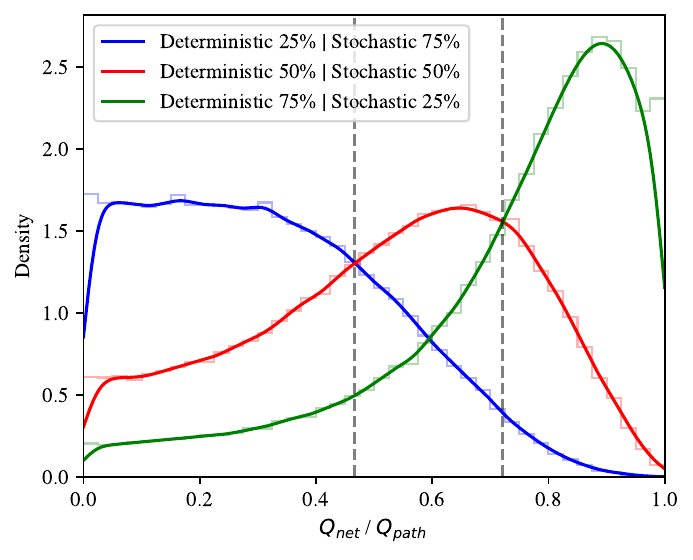}
\caption{A model walk where a fraction of the steps are deterministic and the remainder are stochastic. This figure presents the averaged results of modelling a 100,000 walks under the timing constraints and snapshot spacing of the \textit{in-situ} GC population of m12i. Each colour represents a different partition of steps. The dashed lines represent the boundaries between where different types of walk dominate. The leftmost region corresponds to stochastic evolution dominating, the rightmost region to ordered evolution and the middle region to mixed behaviour.}
\label{fig:k_separation}
\end{figure}

Because not all GCs exist in the host galaxy for the same duration, we run the model over different time frames to match GC formation and accretion histories. For example, in m12i:
\begin{itemize}
    \item \textit{In-situ} GCs form between 1-5 Gyr and so in the model we initialise each particle at a random snapshot in this interval. The results of this iteration of the model are presented in Figure~\ref{fig:k_separation}.
    \item \textit{Ex-situ} GCs are accreted between 2-11 Gyr and so in the model we initialise each particle at a random snapshot in this interval.
\end{itemize}
Between these two instances of the model, the resulting boundaries between different types of motion differ by only $\sim$ 0.01. Therefore we adopt the initial boundaries from the \textit{in-situ} model:
\begin{itemize}
    \item The stochastic dominated boundary: Q$_{\text{net}}$ / Q$_{\text{path}}$ = 0.49
    \item The deterministic dominated boundary: Q$_{\text{net}}$ / Q$_{\text{path}}$ = 0.73
\end{itemize}

For the moving window analysis (10 snapshots per window), we evaluate Q$_{\text{net}}$ / Q$_{\text{path}}$ at each frame. We find that the separation between deterministic and stochastic walks becomes less distinct on this timescale. As such, we instead focus on the boundary between 25–75\% deterministic fractions.

Across most windows, this separation remains relatively constant, with an increase near the final snapshots, where the spacing between snapshots decreases to $\sim$2 Myr. Overall, we find the average 25–75\% separation across all windows to be $\sim$0.65, roughly the average of the 0.49 and 0.73 boundaries determined from the \textit{in-situ} model. Given this approximate match, we retain the 0.49 and 0.73 boundaries in Figure~\ref{fig:m12i_k} to represent the divisions between deterministic and stochastic dominated walks.

\section{Impact of Mergers on Kinematic Evolution}\label{asec:impact_of_mergers}
\begin{figure*}
\includegraphics[width=0.8\textwidth]{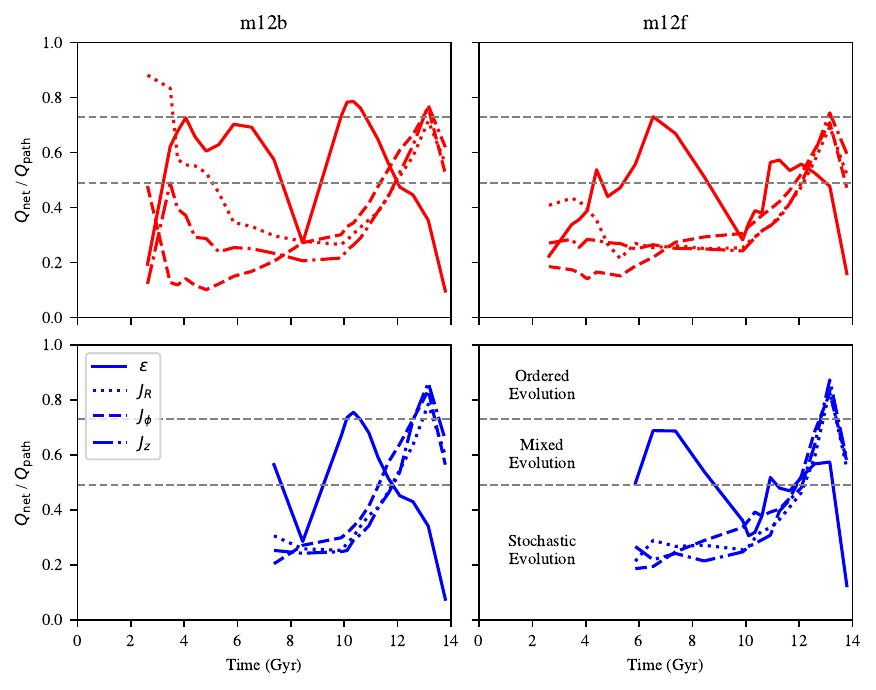}
\caption{Moving window analysis of the ratio Q$_{\text{net}}$/Q${_\text{path}}$, averaged over 100 iterations of the GC model. Each window contains 10 snapshots over which Q$_{\text{net}}$ and Q${_\text{path}}$ are calculated. The left columns is galaxy m12b and the right column is galaxy m12f. The top panel represents the \textit{in-situ} population and the bottom panel is the \textit{ex-situ}. Each panel is divided into three regions, separated by the dashed lines. The upper region corresponds to ordered evolution, the lower region to stochastic evolution and the middle region to mixed behaviour. For visual clarity we have labelled regions in the bottom right panel. Boundary definitions are detailed in Appendix~\ref{asec:random_walk_divisions}. Dips in $\epsilon$ represent the impact of significant mergers. The legend in the bottom-left panel applies to all panels and is independent of colour.}
\label{afig:m12b_m12f_k}
\end{figure*}

Figure~\ref{afig:m12b_m12f_k} is similar to Figure~\ref{fig:m12i_k} for m12i in the main text, but instead illustrate the trends for m12b and m12f, where mergers are sufficiently strong to induce stochastic migration in the normalised energy.


\bsp	
\label{lastpage}
\end{document}